\documentclass[%
 reprint,
 superscriptaddress,
 amsmath,amssymb,
 aps,
prb,
floatfix
]{revtex4-1}

\usepackage[T1]{fontenc}
\usepackage[utf8]{inputenc}
\usepackage[english]{babel}

\usepackage{graphicx}% Include figure files
\usepackage{dcolumn}% Align table columns on decimal point
\usepackage{bm}% bold math
\usepackage[normalem]{ulem}
\usepackage{hyperref}% add hypertext capabilities
\hypersetup{colorlinks = true, citecolor = blue, breaklinks = true}
\graphicspath{{images/}}

\usepackage{textcomp}
\usepackage{multirow}
\usepackage{xfrac}

\usepackage{chemformula}
\setchemformula{kroeger-vink=true}

\usepackage[resetlabels]{multibib}
\newcites{SI}{References}

\begin{document}

\title{Ferroelectricity promoted by cation/anion divacancies in \ch{SrMnO3}}

%list of affiliations to force the order (revtex-4.1 guide)
\affiliation{%
Department of Chemistry, Biochemistry and Pharmaceutical Science, University of Bern, Freiestrasse 3, CH-3012 Bern, Switzerland 
}%
\affiliation{%
National Centre for Computational Design and Discovery of Novel Materials (MARVEL), Switzerland
}%

%now start actual author list
\author{Chiara Ricca}
\affiliation{%
Department of Chemistry, Biochemistry and Pharmaceutical Science, University of Bern, Freiestrasse 3, CH-3012 Bern, Switzerland 
}%
\affiliation{%
National Centre for Computational Design and Discovery of Novel Materials (MARVEL), Switzerland
}%

\author{Danielle Berkowitz}
\affiliation{%
Department of Chemistry, Biochemistry and Pharmaceutical Science, University of Bern, Freiestrasse 3, CH-3012 Bern, Switzerland 
}%

\author{Ulrich Aschauer}
\email{ulrich.aschauer@dcb.unibe.ch}
\affiliation{%
Department of Chemistry, Biochemistry and Pharmaceutical Science, University of Bern, Freiestrasse 3, CH-3012 Bern, Switzerland 
}%
\affiliation{%
National Centre for Computational Design and Discovery of Novel Materials (MARVEL), Switzerland
}%

\date{\today}

\begin{abstract}
We investigate the effect of polar Sr-O vacancy pairs on the electric polarization of \ch{SrMnO3} (SMO) thin films using density functional theory (DFT) calculations. This is motivated by indications that ferroelectricity in complex oxides can be engineered by epitaxial strain but also \textit{via} the defect chemistry. Our results suggest that intrinsic doping by cation and anion divacancies can induce a local polarization in unstrained non-polar SMO thin films and that a ferroelectric state can be stabilized below the critical strain of the stoichiometric material. This polarity is promoted by the electric dipole associated with the defect pair and its coupling to the atomic relaxations upon defect formation that polarize a region around the defect. This suggests that polar defect pairs affect the strain-dependent ferroelectricity in semiconducting antiferromagnetic SMO. For metallic ferromagnetic SMO we find a much weaker coupling between the defect dipole and the polarization due to much stronger electronic screening. Coupling of defect-pair dipoles at high enough concentrations along with their switchable orientation thus makes them a promising route to affect the ferroelectric transition in complex transition metal oxide thin films.
\end{abstract}

\maketitle

\section{\label{sec:intro}Introduction}

Ferroelectricity in complex perovskite oxides has attracted great interest due to  potential applications of ferroelectric thin films for various information storage technologies, such as non volatile random access memories and high-density data storage devices~\cite{Scott954, Lu2012, Liu2014}. Point defects are promising to tailor the functional properties of oxides~\cite{fuchigami2009, tuller2011, kalinin2012, kalinin2013functional, chandrasekaran2013, bivskup2014, bhattacharya2014magnetic, becher2015strain, marthinsen2016coupling, griffin2017defect, rojac2017domain}. In particular, they can affect the polarization response in ferroelectrics by controlling the local polarization and the mechanism and kinetics of polarization switching~\cite{Chu2005, Kalinin2010}. Defect pairs such as cation-anion divacancies or vacancies coupled with substitutional atoms were shown to play an essential role in determining polarization properties~\cite{Yang2013}. For example, \ch{Fe_{Ti}-V_O} defects are able to align in the direction of the lattice polarization in ferroelectric \ch{PbTiO3}~\cite{chandrasekaran2013}. \ch{V_{Pb}-V_{O}} divacancies are an important source of local polarization in Pb-containing perovskite oxides such as \ch{PbTiO3}, where a \ch{V_{Pb}-V_O} concentration of 1.7\% can induce a reduction of the ferroelectric transition temperature by about 35~K~\cite{Chu1995, Cockayne2004}. Finally, defect pairs can also promote ferroelectricity in paraelectric materials: off-centered antisite-like defects consisting of a Sr vacancy and an interstitial Ti atom or by one Ti/Sr antisite defect coupled to an oxygen vacancy or even by Sr-O-O trivacancies are believed to play a pivotal role for emerging room-temperature ferroelectricity in \ch{SrTiO3} thin films~\cite{Choi2009, Kim2009, Yang2015, Klyukin2017}.

\begin{figure}
	\centering
	\includegraphics[width=\columnwidth]{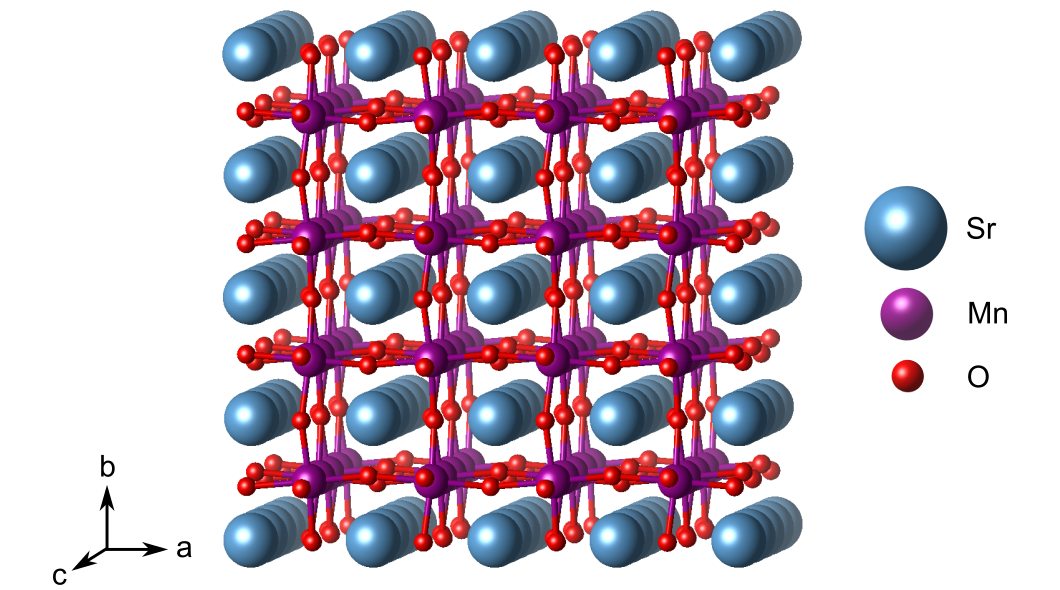}
	\caption{($4\times4\times4$) \textit{Pnma} supercell of stoichiometric \ch{SrMnO3}.}
	\label{fig:SMO_structure_bulk}
\end{figure}
Emerging ferroelectricity in nominally non-polar stoichiometric transition-metal oxides is the result of a complex interplay between structural, electronic, and magnetic degrees of freedom~\cite{SpaldinRenaissance2005, kalinin2013functional}. Biaxial strain, imposed by lattice matching with a substrate during coherent epitaxial thin-film growth can, for instance, stabilize the perovskite phase of \ch{SrMnO3} (SMO, space group $Pnma$, see Fig.~\ref{fig:SMO_structure_bulk}) which has a G-type antiferromagnetic (AFM) order~\cite{chmaissem2001relationship, Kobayashi:2010by}. Moreover, sufficiently large strain can induce a polar distortion in SMO, which involves off-centering of the Mn ions within their oxygen octahedra and leads to ferroelectric behavior~\cite{lee2010epitaxial, becher2015strain}. Theory predicts the in-plane polar modes to soften for tensile strain larger than about~+2.5\%, the polar distortion increasing for larger tensile strain where a transition towards the ferromagnetic (FM) phase is predicted~\cite{lee2010epitaxial}. Compressive strain larger than~5\% induces, instead, ferroelectricity in the direction perpendicular to the strain plane~\cite{marthinsen2016coupling}. As discussed above, the defect chemistry is an additional parameter to consider when designing and controlling ferroelectricity in complex oxides and its coupling or competition with the other degrees of freedom requires careful investigation~\cite{kalinin2013functional}. As such, while tensile strain in SMO promotes ferroelectricity, it also favors oxygen-vacancy formation, the presence of which, in turn, suppresses ferroelectricity~\cite{marthinsen2016coupling}. In SMO, the formation of Sr and O divacancies (\ch{V_{Sr}-V_O}) with an associated defect dipole could be a viable approach to reverse this behavior without recurring to extrinsic doping. Sr deficient \ch{SrMnO_{$3-\delta$}} thin films can be grown by pulsed laser deposition, even if crystallinity is reduced compared to cation balanced or Mn-deficient films~\cite{kobayashi2011cation}.

While the effect of polar defect pairs and strain are thus established separately, their interplay has not been previously considered. In the present work, we investigate - \textit{via} density functional theory (DFT) calculations - the microscopic origin of polarization induced by cation-anion divacancies and gain a deep understanding of the interplay between these polar defect pairs and strain, polarization, electronic properties, structure, and magnetism in SMO thin films. Our results show that the electric dipole pointing from the positively charged \ch{V_{Sr}} to the negatively charged \ch{V_O} indeed induces a lattice polarization around the defect pair. This defect-pair dipole can couple with applied epitaxial strain and induce ferroelectricity for strains below those predicted to stabilize the polar structure in stoichiometric SMO, especially in the G-AFM phase. This interplay strongly depends on the magnetic and electronic properties of the film, the larger electronic screening in the metallic FM phase hindering the coupling between defect-pair dipoles and the polarization in the surrounding crystal.

%%%%%%%%%%%%%%%%%%%%%%%%%%%%%%%%%%%%%%%%%%%%%%%%%%%%%%%%%%%%%%%%%%%%%%%%%%%%%%%%%%%%%%%
%%%%%%%%%%%%%%%%%%%%%% METHODS %%%%%%%%%%%%%%%%%%%%%%%%%%%%%%%%%%%%%%%%%%%%%%%%%%%%%%%%
%%%%%%%%%%%%%%%%%%%%%%%%%%%%%%%%%%%%%%%%%%%%%%%%%%%%%%%%%%%%%%%%%%%%%%%%%%%%%%%%%%%%%%%
\section{Methods}\label{sec:compdetails}

DFT calculations were performed with the {\sc{Quantum ESPRESSO}} package~\cite{giannozzi2009quantum, Giannozzi2017} using PBEsol~\cite{perdew2008pbesol} as exchange-correlation functional and ultrasoft pseudopotentials~\cite{vanderbilt1990soft} with Sr($4s$, $4p$, $5s$), Mn($3p$, $4s$, $3d$), and O($2s$, $2p$) valence states\footnote[3]{Ultrasoft pseudopotentials from the PSLibrary were taken from \url{www.materialscloud.org}: Sr.pbesol-spn-rrkjus\textunderscore psl.1.0.0.UPF, Mn.pbesol-spn-rrkjus\textunderscore psl.0.3.1.UPF, and O.pbesol-n-rrkjus\textunderscore psl.1.0.0.UPF}. Wavefunctions were expanded in plane waves with a kinetic-energy cut-off of 70~Ry and a cut-off of 840~Ry for the augmented density. A Gaussian smearing with a broadening parameter of 0.01 Ry was used in all cases. A Hubbard correction~\cite{anisimov1991band, anisimov1997first, Dudarev1998} was applied on the Mn-$3d$ orbitals within the rotationally invariant formulation of Dudarev~\cite{Dudarev1998} with $U$ values computed self-consistently for the stoichiometric G-AFM and FM SMO phases~\cite{Ricca2019}.

Biaxial epitaxial strain in the \textit{ac}-plane imposed by a cubic substrate was accounted for in the strained-bulk setup as described in Ref.~\cite{Rondinelli:2011jk}. We consider a single relative orientation of the substrate and film, corresponding to strain in the pseudocubic SMO $ac$ plane. Lattice instabilities were calculated at the $\Gamma$ point of a $2\times2\times2$ supercell of the 5-atom primitive $Pm\bar{3}m$ cell using the frozen phonon approach~\cite{Kunc1982} and analyzed using the PHONOPY interface~\cite{TOGO2015}. A shifted $6\times6\times6$ Monkhorst-Pack \textbf{k}-point mesh was used for reciprocal space integration in this case. Defect pairs were calculated in 320-atom $4\times4\times4$ supercells of the ideal 5-atom cubic cell with $\Gamma$-point sampling of the Brillouin zone. We note that, despite the coarser \textbf{k}-mesh compared to the $2\times2\times2$ supercell, a good qualitative description of properties and general trends is retained, while making the computation of these large cells tractable. Before creating defect pairs, all atoms in the 320-atom supercell were displaced along the polar-mode eigenvectors of the stoichiometric structure, the resulting structure being designated as ``prepolarized'' in the following. Defect pairs were created by simultaneously removing one oxygen atom (\ch{V_O}, concentration 0.5\%) and one strontium atom (\ch{V_{Sr}}, concentration 1.6\%) from this supercell. Different relative arrangements of the two vacancies were taken into account (see Sec.~\ref{sec:ef_confs}). Since in fully or partially ionic compounds it is generally favorable for vacancies to be charge balanced by other defects, only the charge neutral Schottky defect pair was taken into account (\ch{V_{Sr}^{''}-V_{O}^{..}} in Kr{\"o}ger-Vink notation~\cite{KROGER1956307}, where the prime and dot symbols indicate, respectively, a charge of -1 and +1 relative to the respective lattice site). For simplicity, we will refer to these defect pairs as \ch{V_{Sr}-V_{O}}. For defective cells, only atomic positions were relaxed with the lattice vectors fixed at the optimized values of the corresponding stoichiometric cell. Convergence thresholds of $1.4\times10^{-5}$~eV for the energy and $5\times10^{-2}$~eV/\AA\ for the forces are used for all relaxations.

The strain-dependent \ch{V_{Sr}-V_{O}} formation energy ($E_{\textrm{f}}$) was computed according to Ref.~\cite{freysoldt2014first}:
\begin{align}
	E_{\textrm{f}}(\epsilon, \mu_\textrm{O},\mu_\textrm{Sr})= E_{\textrm{def}}(\epsilon) -E_{\textrm{stoi}}(\epsilon) + \mu_\textrm{O} + \mu_\textrm{Sr}\,,
\end{align}
where $E_{\textrm{def}}$ and $E_{\textrm{stoi}}$ are the DFT total energies of the defective and stoichiometric cell, respectively, $\epsilon$ is the applied strain, and $\mu_\textrm{O}$ and $\mu_\textrm{Sr}$ are the O and Sr chemical potential, respectively. We will report results in the O-poor limit, \textit{i.e.} $\mu_\textrm{O} = \frac{1}{2}\textrm{E}(\textrm{O}_2)+ \Delta \mu_{\textrm{O}}$ with $\textrm{E}(\textrm{O}_2)$ being the energy of an oxygen molecule and $\Delta \mu_{\textrm{O}} =$ -1.39~eV. The Sr chemical potential ($\mu_\textrm{Sr} = \textrm{E}_\textrm{Sr}+ \Delta \mu_{\textrm{Sr}}$) was derived as function of $\mu_\textrm{O}$, $\textrm{E}_\textrm{Sr}$ being the total energy of metallic Sr and $\Delta \mu_{\textrm{Sr}}=$-4.28~eV under O-poor conditions. The above limit to the O chemical potential was derived considering the stability of the system ($\Delta \mu_{\textrm{Sr}} + \Delta \mu_{\textrm{Mn}} + 3\Delta \mu_{\textrm{O}} = \Delta H_f(\mathrm{SMO})=-10.22$ eV) against decomposition to elemental Sr ($\Delta \mu_{\textrm{Sr}} \le 0$) and Mn ($\Delta \mu_{\textrm{Mn}}\le 0$) and against SrO ($ \Delta \mu_{\textrm{Sr}} +  \Delta \mu_{\textrm{O}} \le \Delta H_f(\mathrm{SrO})=-5.67$ eV) and MnO ($\Delta \mu_{\textrm{Mn}} + \Delta \mu_{\textrm{O}} \le \Delta H_f(\mathrm{MnO})=-3.16$ eV) formation, where $\Delta H_f$ indicates the computed heats of formation. $\Delta H_f$ for transition-metal oxides were corrected according to Ref.~\cite{ceder20011} to account for mixing of DFT and DFT+$U$ total energies.

The polarization $\vec{P}$ was estimated using a point-charge model:
\begin{equation}
	\vec{P}=\sum_i \vec{r}_i q_i\,,
	\label{eq:polarization1}
\end{equation}
where $\vec{r}_i$ is the position of atom $i$ and $q_i$ is its formal charge: +2 for Sr, -2 for O, and +4 for Mn. The polarization, being a multivalued quantity~\cite{Spaldin:2012ts}, has been corrected by an integer number of polarization quanta $\vec{Q}$, computed as:
\begin{equation}
	\vec{Q}=\frac{e}{V} \begin{bmatrix} a\\b\\c \end{bmatrix}\,,
	\label{eq:polarization2}
\end{equation}
with $a$, $b$, and $c$ being the lattice parameters, $V$ the volume of the unit cell, and $e$ the elementary charge. This model includes both the lattice contribution and the effect of the electric dipole associated with the defect pair from the positively charged \ch{V_{Sr}^{''}} to the negatively charged \ch{V_O^{..}}, but neglects the electronic contribution to the ferroelectric polarization compared to other approaches such as the Berry phase formalism~\cite{berryphase1, berryphase2}. It was nonetheless adopted both to reduce the computational cost and to allow comparison of the behavior of the G-AFM and FM phases, the Berry phase method not being applicable to metallic systems such as FM SMO.

Barriers for polarization switching were calculated using the climbing-image nudged elastic band (CI-NEB) method~\cite{Henkelman2000}. Minimum energy pathways were relaxed until forces on each image converged below $1\times10^{-3}$~eV/\AA.

%%%%%%%%%%%%%%%%%%%%%%%%%%%%%%%%%%%%%%%%%%%%%%%%%%%%%%%%%%%%%%%%%%%%%%%%%%%%%%%%%%%%%%%
%%%%%%%%%%%%%%%%%%%%% RESULTS %%%%%%%%%%%%%%%%%%%%%%%%%%%%%%%%%%%%%%%%%%%%%%%%%%%%%%%%%
%%%%%%%%%%%%%%%%%%%%%%%%%%%%%%%%%%%%%%%%%%%%%%%%%%%%%%%%%%%%%%%%%%%%%%%%%%%%%%%%%%%%%%%
\section{Results and Discussion\label{sec:results}}

%%%%%%%%%%%%%%%%%%%%%%%%%%%%%%%%%%%%%%%%%%%%%%%%%%%%%%%%%%%%%%%%%%%%%%%%%%%%%%%%%%%%%%%
\subsection{Interplay of strain, magnetism, and ferroelectricity in stoichiometric SMO}
%%%%%%%%%%%%%%%%%%%%%%%%%%%%%%%%%%%%%%%%%%%%%%%%%%%%%%%%%%%%%%%%%%%%%%%%%%%%%%%%%%%%%%%

Before turning to polar \ch{V_{Sr}-V_{O}} defect pairs, we investigate the interplay between strain, magnetism, and ferroelectricity in stoichiometric SMO, which will be fundamental to understand the interplay between strain and the defect-induced properties. Strain-dependent ferroelectricity in stoichiometric G-AFM SMO thin films was previously reported by Marthinsen \textit{et al.}~\cite{marthinsen2016coupling}. Here, we evaluate polar instabilities also for the FM phase but note that its metallicity will preclude ferroelectric switching, an unstable polar mode indicating merely a polar metal state \cite{zhou2020review, benedek2016ferroelectric}.

\begin{figure}
	\centering
	\includegraphics[width=\columnwidth]{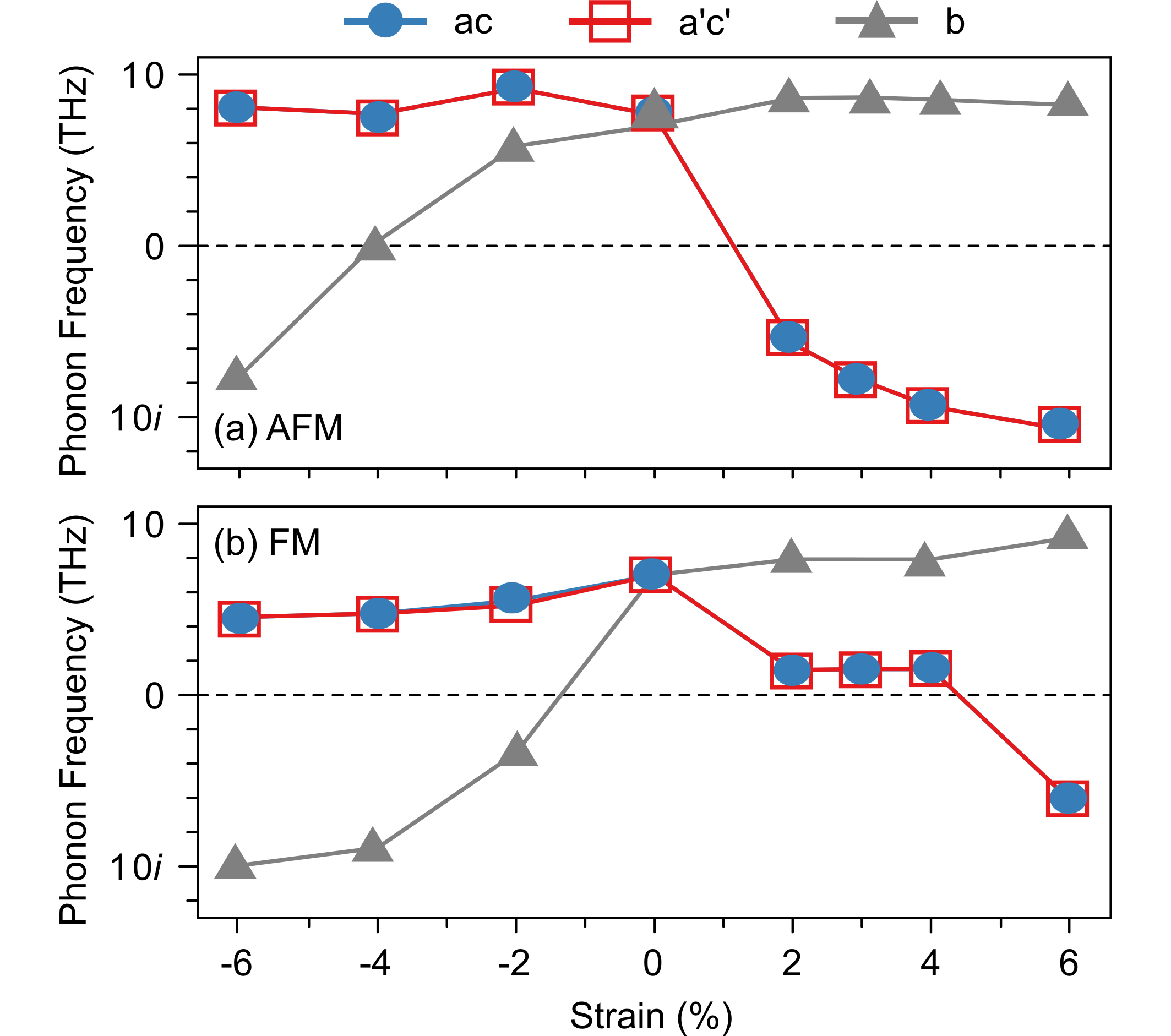}
	\caption{Evolution of the polar-mode phonon frequencies as a function of biaxial epitaxial strain in the $Pnma$ SMO structure for a) G-AFM and b) FM order.}
	\label{fig:Phonons}
\end{figure}
Fig.~\ref{fig:Phonons} shows the evolution of SMO polar-mode frequencies as a function of strain. In unstrained SMO the G-AFM phase is dynamically stable (see Fig.~\ref{fig:Phonons}a). The double-degenerate in-plane (IP) polar modes, associated with the displacement of Mn atoms from the center of their oxygen octahedra in the $ac$-plane, become unstable at about 2\% tensile strain, while the out-of plane (OP) mode softens between 4 and 6\% compressive strain. These results are in excellent agreement with Ref.~\cite{marthinsen2016coupling}, differences in the critical strain being attributed to different Hubbard $U$ values~\cite{Hong2012}. The FM phase exhibits a different strain-dependence: not only are the modes generally softer than in the AFM phase, but, more importantly, the IP modes become unstable only for large tensile strain of about 6\%, while the OP mode softens already at -2\%. As we will show in the following, this different behavior will affect the effect of polar defect pairs on the ferroelectric behavior of SMO thin films.

%%%%%%%%%%%%%%%%%%%%%%%%%%%%%%%%%%%%%%%%%%%%%%%%%%%%%%%%%%%%%%%%%%%%%%%%%%%%%%%%%%%%%%%
\subsection{\ch{V_{Sr}-V_{O}} formation energy and relative stability\label{sec:ef_confs}}
%%%%%%%%%%%%%%%%%%%%%%%%%%%%%%%%%%%%%%%%%%%%%%%%%%%%%%%%%%%%%%%%%%%%%%%%%%%%%%%%%%%%%%%

%
\begin{figure}
 \centering
 \includegraphics[width=\columnwidth]{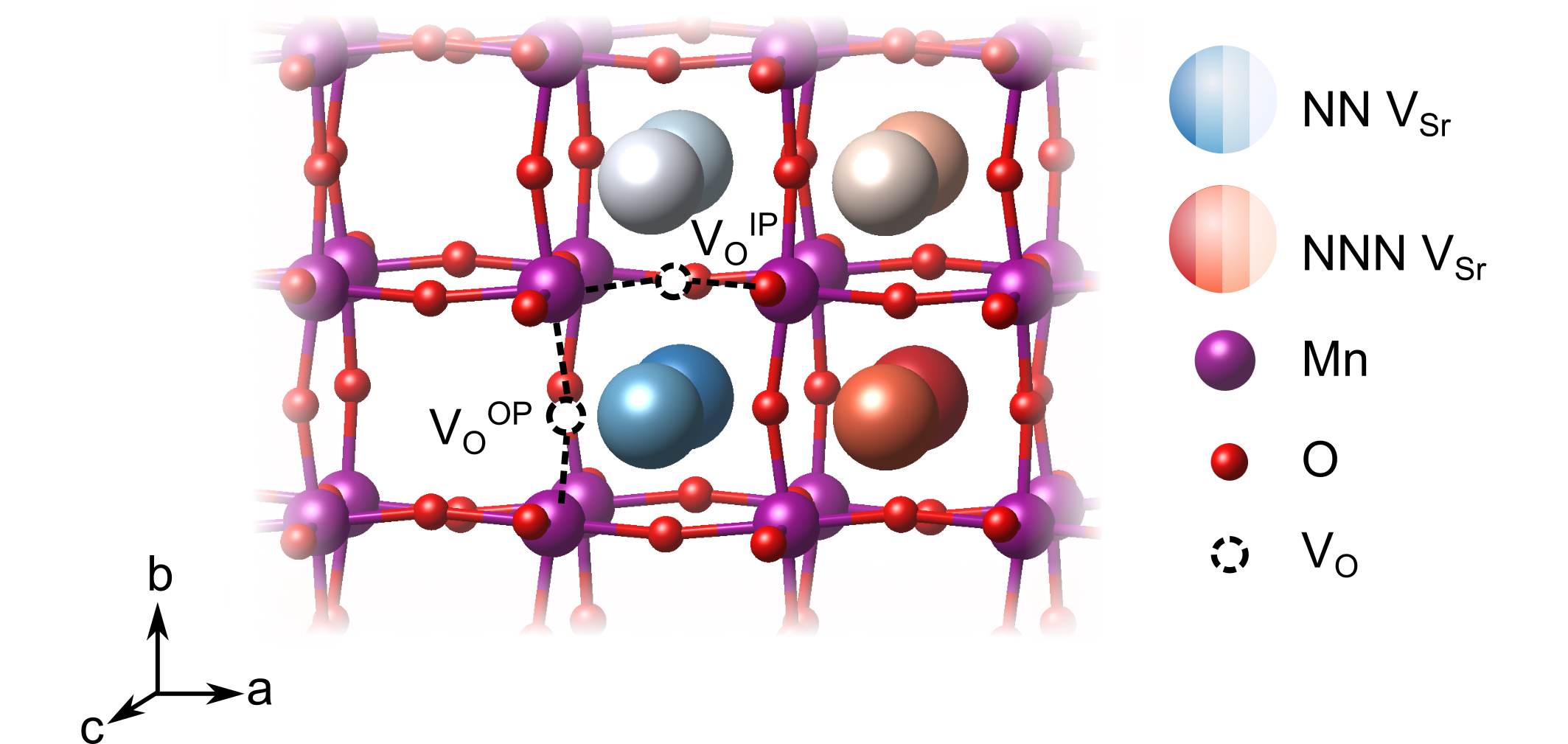}
 \caption{Schematic representation of the possible relative arrangements of \ch{V_{Sr}-V_{O}} defect pairs in the 320-atom SMO cell for an out-of-plane (OP) or in-plane (IP) oxygen vacancy. The different shades of blue refer to configurations in which \ch{V_{Sr}} is nearest-neighbor (NN) to \ch{V_O}, while the Sr sites in different shades of red are for \ch{V_{Sr}} in next-nearest neighbor (NNN) positions.}
\label{fig:confs}
\end{figure}
We have studied \ch{V_{Sr}-V_{O}} defects in a large 320-atom $4\times4\times4$ SMO supercell to allow isolating the effect of individual defect dipoles. In smaller cells, the interaction of the defect and image dipoles overestimates relaxation energies relative to an isolated defect~\cite{Cockayne2004}. As can be seen in Fig.~\ref{fig:confs}, there are two symmetry-distinct oxygen-vacancy positions: an in-plane (IP) and and out-of-plane (OP) O atom with the broken \ch{Mn-O-Mn} bond respectively in the biaxial strain ($ac$) plane and perpendicular to it. For each \ch{V_O}, we tested inequivalent sites for a \ch{V_{Sr}} in nearest-neighbor (NN) or next-nearest neighbor (NNN) positions to the oxygen vacancy. When a \ch{V_O^{IP}} with the broken \ch{Mn-O-Mn} bond along the $a$-axis is created, the four \ch{V_{Sr}^{NN}} positions indicated in different shades of blue in Fig.~\ref{fig:confs} correspond to all possible orientations of the \ch{V_{Sr}^{''}-V_O^{..}} dipole in the $bc$ plane. Instead, for the \ch{V_{Sr}^{NNN}} indicated by different shades of red in Fig.~\ref{fig:confs}, one could also identify four additional equivalent configurations with \ch{V_{Sr}^{NNN}} located at negative $a$ coordinates with respect to the \ch{V_O}. Similar arguments apply for \ch{V_{Sr}} in NN and NNN positions with respect to \ch{V_O^{OP}}. For simplicity and to reduce the computational cost, we considered only the 16 \ch{V_{Sr}-V_{O}} configurations indicated in Fig.~\ref{fig:confs}.

\begin{figure}
 \centering
 \includegraphics[width=\columnwidth]{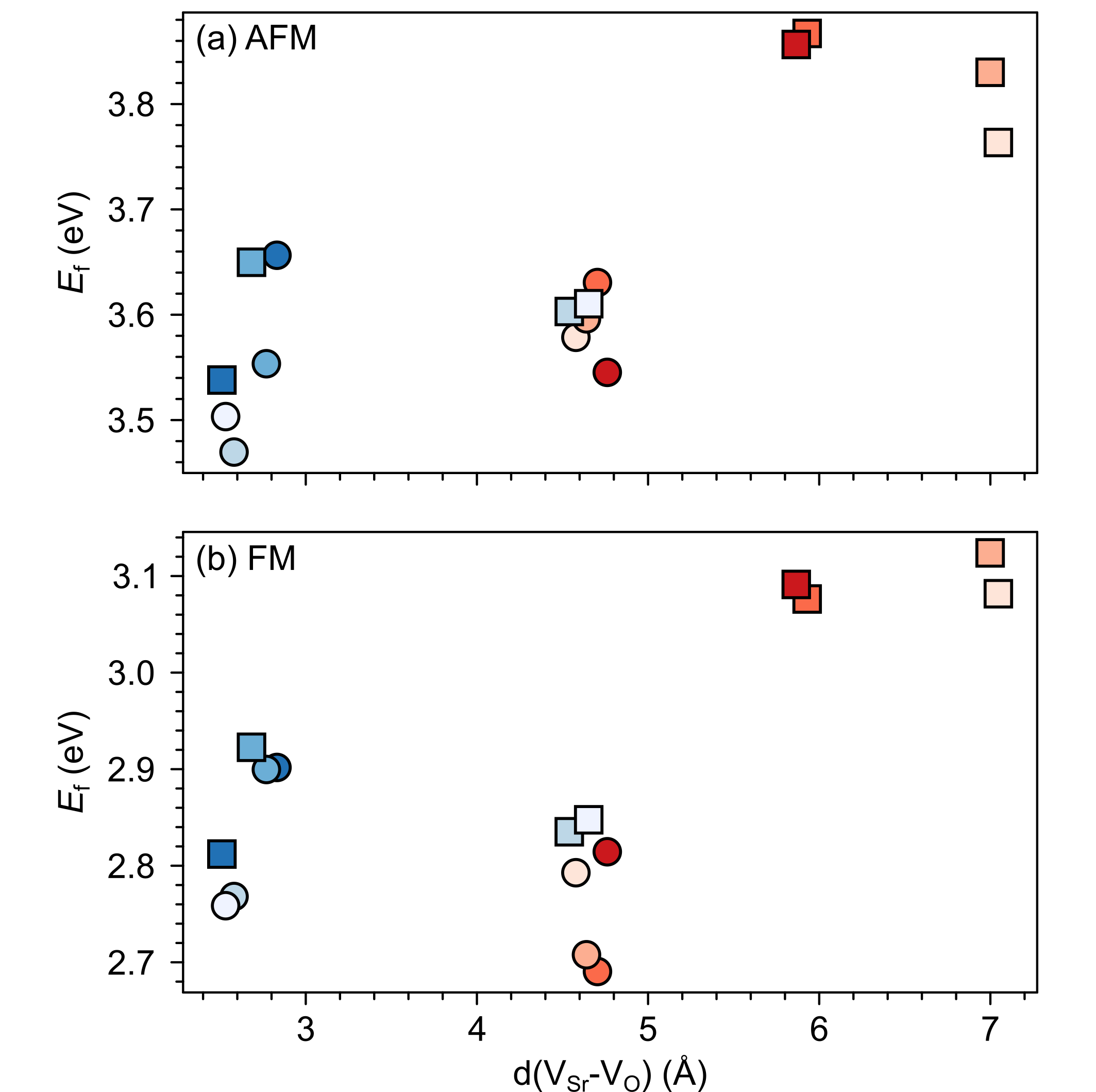}
 \caption{Formation energy ($E_\mathrm{f}$) for \ch{V_{Sr}-V_{O}} defect pairs in unstrained (a) AFM and (b) FM SMO as a function of the distance between \ch{V_{Sr}} and \ch{V_O}. Circle and square symbols refer to \ch{V_O^{IP}} and \ch{V_O^{OP}}, respectively. See Fig.~\ref{fig:confs} for the color code.}
\label{fig:Eform_unstrained}
\end{figure}
We first investigate the relative stability of the different \ch{V_{Sr}-V_{O}} configurations in unstrained SMO characterised by their formation energy in Fig.~\ref{fig:Eform_unstrained}. Divacancies with a \ch{V_O^{IP}} (circles) are usually more stable than those with a \ch{V_O^{OP}} (squares). Generally, \ch{V_{Sr}} prefer to be close to the \ch{V_O}: the most stable stable defect pairs are \ch{V_{Sr}^{NN}-V_O^{IP}}, followed by configurations with a \ch{V_{Sr}^{NNN}} at about 4.9~\AA\ from the oxygen vacancy that have slightly (0.1~eV) larger formation energies. \ch{V_{Sr}^{NNN}-V_O^{OP}} defects, where the two vacancies are separated by as much as 6-7\AA\ and interact less, are the least likely to form. Compared to the semiconducting AFM phase, the metallic nature of the FM order results in smaller energetic differences between the configurations, as well as generally in lower formation energies.

\begin{figure}
 \centering
 \includegraphics[width=\columnwidth]{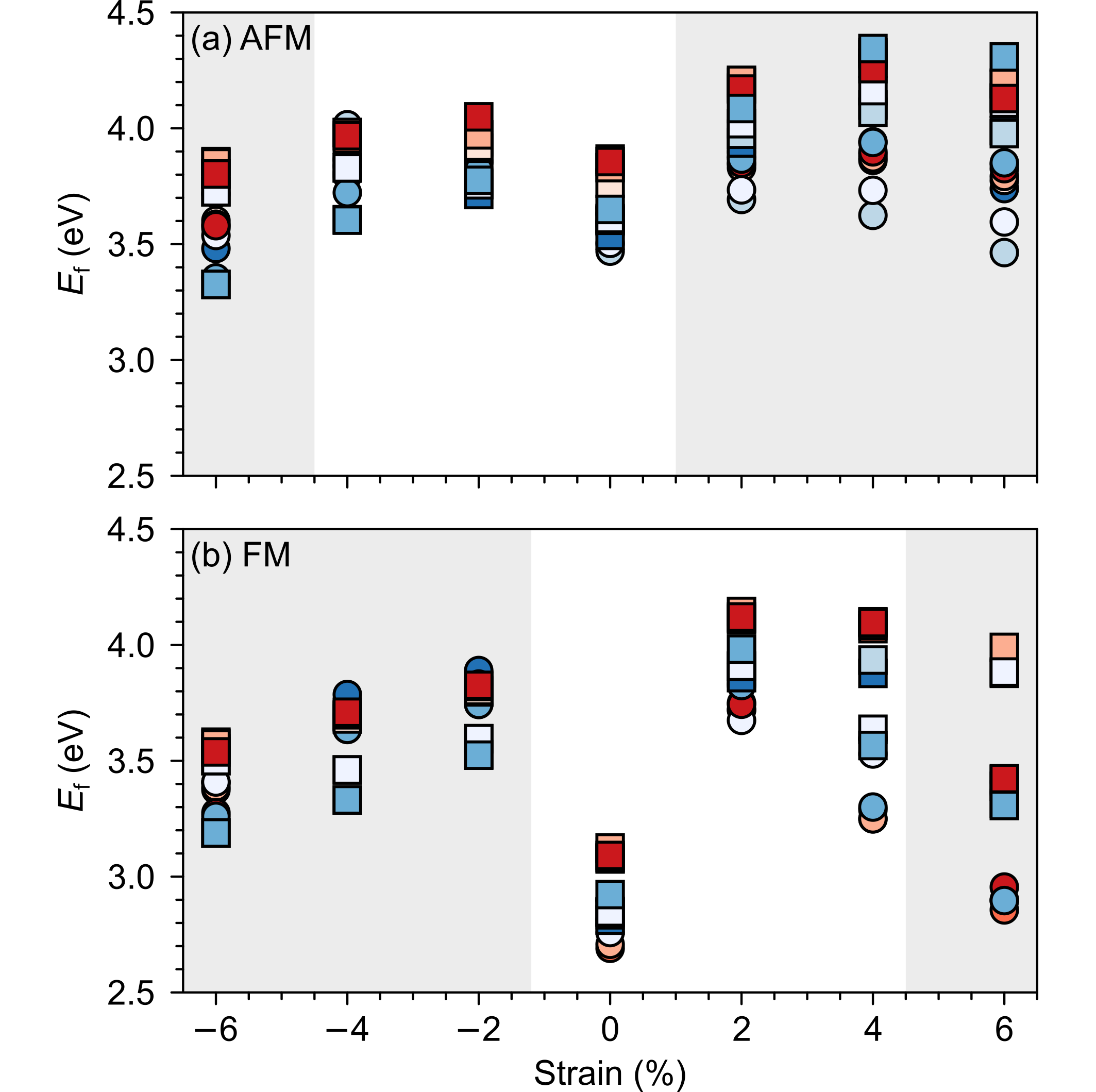}
 \caption{Formation energy ($E_\mathrm{f}$) for \ch{V_{Sr}-V_{O}} defects in (a) AFM and (b) FM SMO as a function of biaxial strain. The shaded grey areas indicate strain ranges with unstable polar modes in stoichiometric SMO. Circle and square symbols refer to \ch{V_O^{IP}} and \ch{V_O^{OP}}, respectively. See Fig.~\ref{fig:confs} for the color code.}
\label{fig:Eform_strain}
\end{figure}
Changes in SMO structural and electronic properties induced by epitaxial strain were shown to influence the \ch{V_O} formation and ordering at inequivalent sites~\cite{aschauer2013strain, aschauer2016interplay, marthinsen2016coupling}. Therefore, it is important to understand the interplay between defect chemistry, strain and magnetism also for \ch{V_{Sr}-V_{O}} defect pairs. In the AFM phase, the defect formation energy ($E_\mathrm{f}$, see Fig.~\ref{fig:Eform_strain}a) exhibits a non-monotonic strain dependence with changes in the sign of the slope close to the critical strains for the ferroelectric instability (Fig.~\ref{fig:Phonons}a). In particular, $E_\mathrm{f}$ increases going from 0\% to -4\% or to +2\% strain, where no polar instability exists. For larger tensile strain, when the IP polar modes become unstable, $E_\mathrm{f}$ increases or stays constant for defect pairs with a \ch{V_O^{OP}}, but decreases especially for \ch{V_{Sr}^{NN}-V_O^{IP}}, which allows for strain-controlled defect ordering. Similarly, at 6\% compressive strain, the formation energy is slightly reduced but a mixture of \ch{V_O^{IP}} and \ch{V_O^{IP}} defects is formed. In the FM phase, instead, $E_\mathrm{f}$ strongly increases going from the unstrained structure to the $\pm$2\% strained geometries and then decreases for larger strain, tensile and compressive strain strongly favoring defect pairs with \ch{V_O^{IP}} and \ch{V_O^{OP}}, respectively. 

As we will further verify in the following sections, these results suggest a stronger coupling between the polar defect pair and the ferroelectric degrees of freedom in the AFM phase, compared to the metallic FM phase, where a stronger electronic screening of the defect dipole occurs. 

%%%%%%%%%%%%%%%%%%%%%%%%%%%%%%%%%%%%%%%%%%%%%%%%%%%%%%%%%%%%%%%%%%%%%%%%%%%%%%%%%%%%%%%
\subsection{Magnetic Order}
%%%%%%%%%%%%%%%%%%%%%%%%%%%%%%%%%%%%%%%%%%%%%%%%%%%%%%%%%%%%%%%%%%%%%%%%%%%%%%%%%%%%%%%

G-AFM is the ground state for stoichiometric bulk SMO but strain and/or defects can induce transition towards the FM phase. For example, in our previous work~\cite{Ricca2019}, using the same computational setup, we observed that 2\% tensile strain can stabilize the FM order in stoichiometric SMO, while a concentration of 4.2\% of oxygen vacancies leads to a FM ground state already for unstrained SMO. For oxygen vacancies, the magnetic transition is generally rationalized by \ch{Mn^{4+}-Mn^{3+}} double exchange due to reduced \ch{Mn^{3+}} sites upon \ch{V_O} formation.

\begin{figure}
 \centering
 \includegraphics[width=\columnwidth]{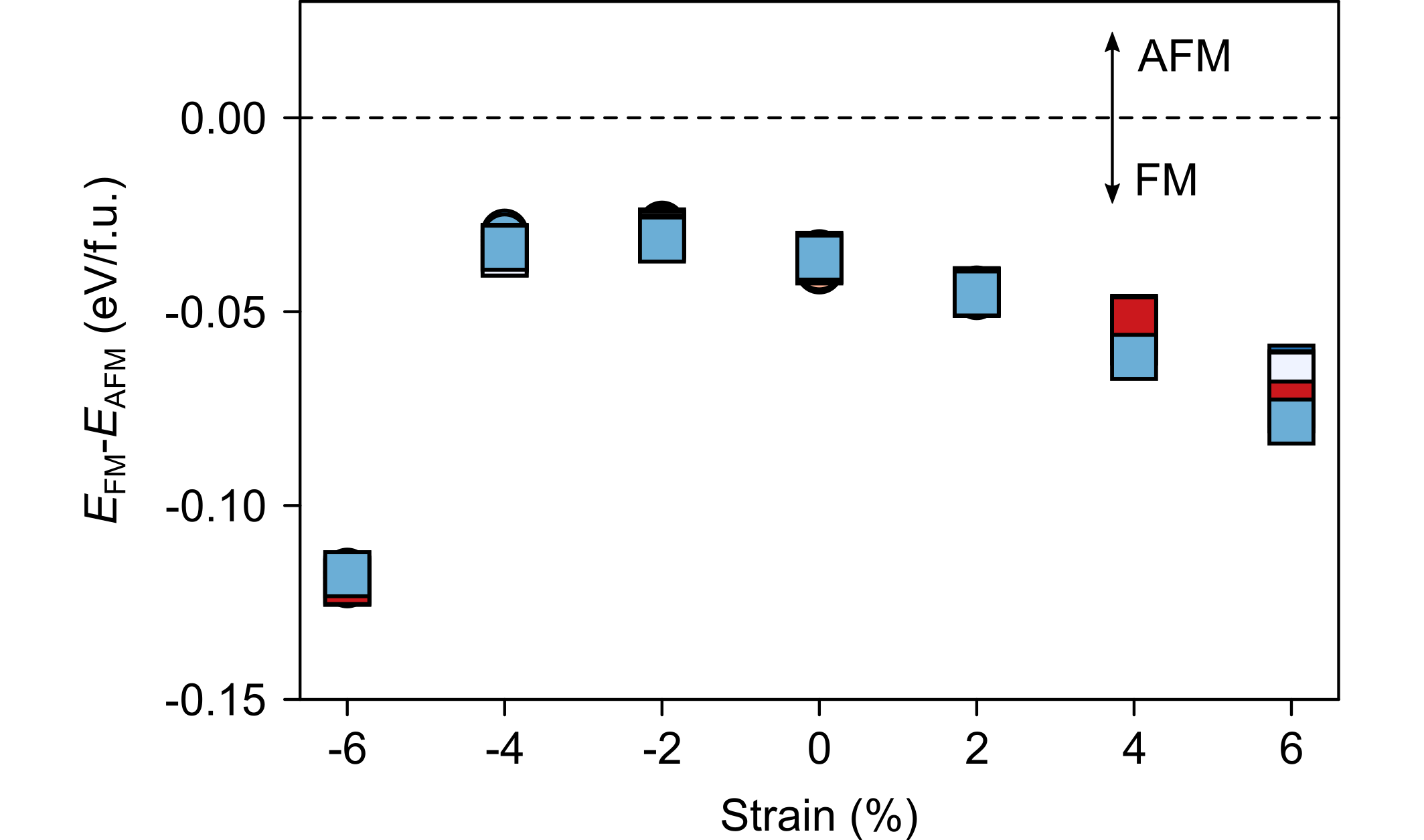}
 \caption{Total energy difference $E_\mathrm{FM}-E_\mathrm{AFM}$ per formula unit between defective cells with FM and AFM magnetic order as a function of the applied epitaxial strain. AFM is more stable for positive and FM for negative differences. Circle and square symbols refer to \ch{V_O^{IP}} and \ch{V_O^{OP}}, respectively. See Fig.~\ref{fig:confs} for the color code.}
\label{fig:magneticorder_strain}
\end{figure}

Fig.~\ref{fig:magneticorder_strain} suggests that, in the considered strain range, \ch{V_{Sr}-V_{O}} defect pairs very slightly favor the FM phase, the preference for the ferromagnetic order increasing under tensile strain. The strong stabilization of the FM phase for -6\% strain can be explained by the strong band-gap reduction in the AFM phase for such large compressive strain (cf. SI\dag\ Fig.~\ref{fig:eg_vs_strain}). Nevertheless the preference for the FM order is difficult to rationalize since charge compensated neutral \ch{V_{Sr}-V_{O}} defect pairs should not lead to reduced \ch{Mn^{3+}}, which is responsible for the emergence of FM order. However, for \ch{V_{Sr}-V_{O}} defect pairs in the AFM phase we observed one or two partially reduced Mn sites (\ch{Mn^{($3+\delta$)+}}), as can be seen from the density of states reported for one \ch{V_O^{IP}} and one \ch{V_O^{OP}} in SI\dag\ Fig.~\ref{fig:PDOS}. We believe these \ch{Mn^{($3+\delta$)+}}, and in turn the predicted stabilization of the FM order, to be a consequence of the interplay between structural relaxations taking place upon \ch{V_O} formation and the established underestimation of SMO band-gap within DFT+$U$: the elongation of \ch{Mn-O} bonds upon \ch{V_O} formation results in the stabilization of the corresponding $e_g$ orbitals, the energy of which, due to the underestimation of the band gap, is lowered to just below the Fermi energy. This results in a partial occupation of this state, the observed \ch{Mn^{($3+\delta$)+}} and the preference for the FM order. For this reason, we caution against the conclusion that \ch{V_{Sr}-V_{O}} defect pairs favor FM and also compute the polarization in defective SMO in the following section by assuming that no Mn reduction takes place upon \ch{V_{Sr}-V_{O}} formation. 
%

%%%%%%%%%%%%%%%%%%%%%%%%%%%%%%%%%%%%%%%%%%%%%%%%%%%%%%%%%%%%%%%%%%%%%%%%%%%%%%%%%%%%%%%
\subsection{Polarization}
%%%%%%%%%%%%%%%%%%%%%%%%%%%%%%%%%%%%%%%%%%%%%%%%%%%%%%%%%%%%%%%%%%%%%%%%%%%%%%%%%%%%%%%

%%%%%%%%%%%%%%%%%%%%%%%%%%%%%%%%%%%%%%%%%%%%%%%%%%%%%%%%%%%%%%%%%%%%%%%%%%%%%%%%%%%%%%%
\subsubsection{Interplay between polar defects and structural relaxations}
%%%%%%%%%%%%%%%%%%%%%%%%%%%%%%%%%%%%%%%%%%%%%%%%%%%%%%%%%%%%%%%%%%%%%%%%%%%%%%%%%%%%%%%
%
\begin{figure}[t]
 \centering
 \includegraphics[width=\columnwidth]{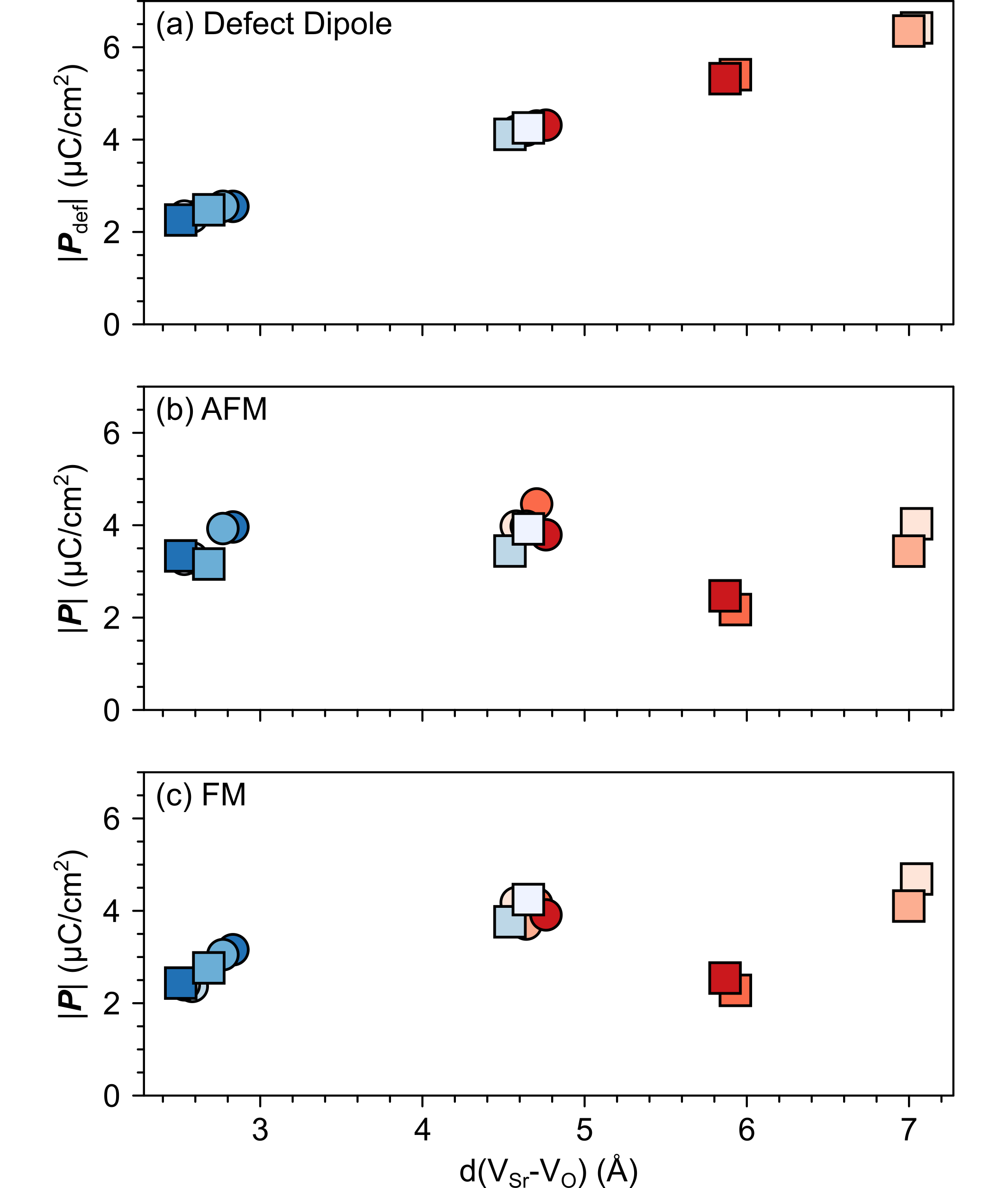}
 \caption{Magnitude of the total polarization vector as a function of the \ch{V_{Sr}-V_{O}} distance computed for different \ch{V_{Sr}-V_{O}} configurations considering a) only the contribution of the defect pair (Eq.~\ref{eq:polarization3}) or b) and c) also lattice contributions (Eq.~\ref{eq:polarization1}) for the AFM and FM phases respectively. Circle and square symbols refer to data obtained for \ch{V_O^{IP}} and \ch{V_O^{OP}}, respectively. See Fig.~\ref{fig:confs} for the color code.}
\label{fig:polarization_defect_dipole}
\end{figure}

Polarization in a nominally non-polar SMO thin film with \ch{V_{Sr}-V_{O}} can arise due to the electric defect dipole ($\vec{D}$) from the negatively charged \ch{V_{Sr}^{''}} to the positively charged \ch{V_{O}^{..}}, which results in the charge center being offset from the geometric center of the cell~\cite{Wang2017nano}. Within a very simple ionic model, the polarization induced by the vacancy pair can be estimated as:
\begin{align}
\vec{P}_\mathrm{def} = 2e \vec{r}_{\ch{V_{Sr}-V_{O}}}/V, \label{eq:polarization3}
\end{align}
where $e$ is the elementary charge, $V$ the cell volume, and $\vec{r}_{\ch{V_{Sr}-V_{O}}}$ is the separation vector between the \ch{V_{Sr}} and \ch{V_O} sites~\cite{Cockayne2004}. The polarization ($\vec{P}_\mathrm{def}$) predicted with this simple model increases linearly with the distance between the two vacancies as shown in Fig.~\ref{fig:polarization_defect_dipole}a. However, when considering the lattice polarization \textit{via} Eq.~\ref{eq:polarization1}, for both AFM and FM phases (Fig.~\ref{fig:polarization_defect_dipole}b and c respectively) \ch{V_{Sr}^{NNN}}-\ch{V_O^{OP}} configurations, characterized by the largest \ch{V_{Sr}-V_{O}} separation, have a polarization much smaller than $\vec{P}_\mathrm{def}$. This suggests that, when \ch{V_{Sr}} and \ch{V_O} are separated by more than 5~\AA, defect-defect interactions are screened, which explains the formation energies in Fig.~\ref{fig:Eform_unstrained}. Similar magnitudes of $\vec{P}$ and $\vec{P}_\mathrm{def}$ are obtained for the remaining \ch{V_{Sr}^{NNN}}-\ch{V_O} configurations, with polarizations larger than $\vec{P}_\mathrm{def}$ when the two defects are in NN positions. This is especially the case for the most stable \ch{V_{Sr}^{NN}-V_O^{IP}} defects in the AFM phase, where a polarization almost twice as large as $\vec{P}_\mathrm{def}$ is obtained (cf. Fig.~\ref{fig:polarization_defect_dipole}a and b).

\begin{figure}
 \centering
 \includegraphics[width=\columnwidth]{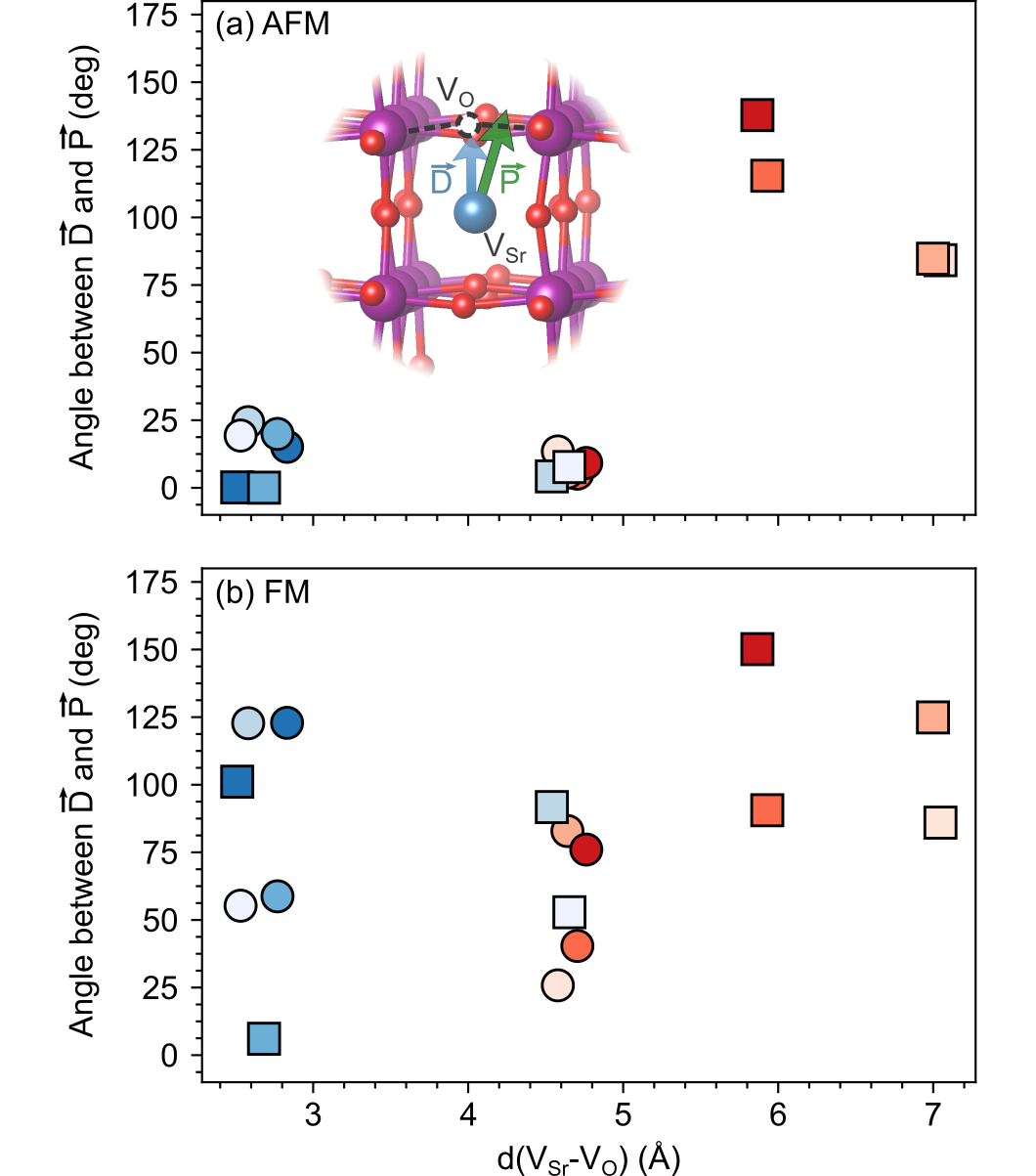}
 \caption{Angle between the defect dipole ($\vec{D}$) and the polarization ($\vec{P}$) computed as a function of the \ch{V_{Sr}-V_{O}} separation for unstrained a) AFM and b) FM SMO. Circle and square symbols refer to \ch{V_O^{IP}} and \ch{V_O^{OP}}, respectively. See Fig.~\ref{fig:confs} for the color code.}
\label{fig:angle_p_d}
\end{figure}
The larger coupling between polar defect pairs and the polarization in the AFM phase is reflected by the angles between the polarization vector $\vec{P}$ and the defect-dipole vector $\vec{D}$, which are quite small (lower than 30$^{\circ}$) in the AFM phase for \ch{V_{Sr}-V_{O}} separated by less than 5~\AA\ (see Fig.~\ref{fig:angle_p_d}). This suggests an alignment of the polarization with the defect dipole in these cases. Larger angles are observed for cells containing \ch{V_{Sr}^{NNN}-V_O^{OP}} at larger separation. In the FM phase (see Fig.~\ref{fig:angle_p_d}b) the angles are also larger due to enhanced electronic screening in this metallic phase that prevents the strong coupling between the defect dipole and polar displacements like in the semiconducting AFM phase.

\begin{figure}[t]
 \centering
 \includegraphics[width=\columnwidth]{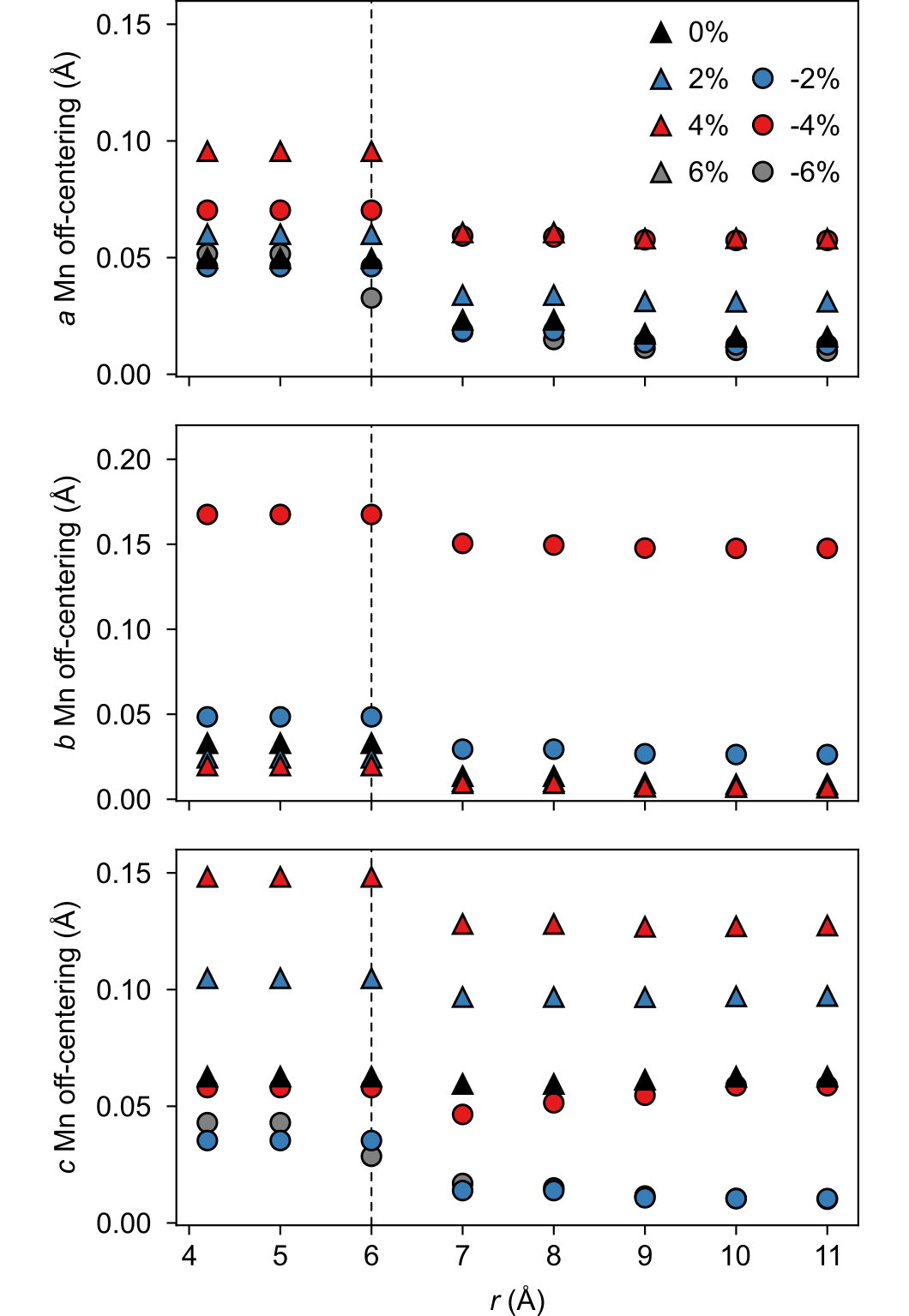}
 \caption{Total Mn off-centering resolved along the a) $a$- , b) $b$-, and c) $c$-axis for Mn atoms lying within a sphere of radius $r$ centered on the \ch{V_{Sr}} position. Results for different amounts of strain are reported for the \ch{V_{Sr}-V_{O}} configuration that is most stable at 0\% strain.}
\label{fig:Mnoffcent_size_Sr}
\end{figure}
The polarization in the defective AFM cell can, indeed, be explained by the atomic displacements upon \ch{V_{Sr}-V_{O}} defect-pair formation. Large displacements from the high-symmetry positions and mainly for Mn atoms in the neighborhood of the \ch{V_{Sr}} take place in the AFM phase. These Mn atoms move towards the cation vacancy, except for sites adjacent to the \ch{V_O} that are more strongly affected by \ch{Mn-O-Mn} bond breaking (see SI\dag\ Fig.~\ref{fig:Mnrelaxations}a,d). This suggests that the larger polarization arises due to the defect pair inducing polar distortions in the surrounding octahedra. We initially focus on the unstrained structure shown by black triangles in Fig.~\ref{fig:Mnoffcent_size_Sr} and will discuss the strain dependence in the next subsection. As shown in Fig.~\ref{fig:Mnoffcent_size_Sr} a larger Mn off-centering is observed in a sphere of about 6~\AA\ around \ch{V_{Sr}} compared to Mn ions further from the cation vacancy. When the two defects are separated by more than 6~\AA\ and do not interact, as for \ch{V_{Sr}^{NNN}-V_O^{OP}} defects, the lattice contraction around the cation vacancy dominates and explains the smaller coupling between the defect and the lattice polarization (see SI\dag\ Fig.~\ref{fig:Mnrelaxations}b,c). Finally, structural relaxations can also explain why, even when strongly interacting, $\vec{P}$ is not perfectly aligned with $\vec{D}$: the small angle between the two vectors stems from displacements of the Mn atoms in NNN positions to \ch{V_O} along the axis of the broken \ch{Mn-O-Mn} (see SI\dag\ Sec.~\ref{sec:si_Mnrelaxations} for more details).

%%%%%%%%%%%%%%%%%%%%%%%%%%%%%%%%%%%%%%%%%%%%%%%%%%%%%%%%%%%%%%%%%%%%%%%%%%%%%%%%%%%%%%%
\subsubsection{Interplay between the polar defect, strain, and polarization}
%%%%%%%%%%%%%%%%%%%%%%%%%%%%%%%%%%%%%%%%%%%%%%%%%%%%%%%%%%%%%%%%%%%%%%%%%%%%%%%%%%%%%%%

The above results indicate that both polar defect pairs and strain can be used to engineer polarity/ferroelectricity in non-polar complex oxides. The effect of \ch{V_{Sr}-V_{O}} divacancies is however local and does not, by itself, lead to a ferroelectric phase. Due to the alignment between the defect dipole and the local polarization, it seems, however, likely that polar defect pairs could help to induce the ferroelectric phase a smaller  strains than in the stoichiometric material. In this section, we will therefore, investigate how the interaction between epitaxial strain and the defect chemistry, influences the polarization of SMO thin films with the two investigated magnetic orders. 

\begin{figure}
 \centering
 \includegraphics[width=\columnwidth]{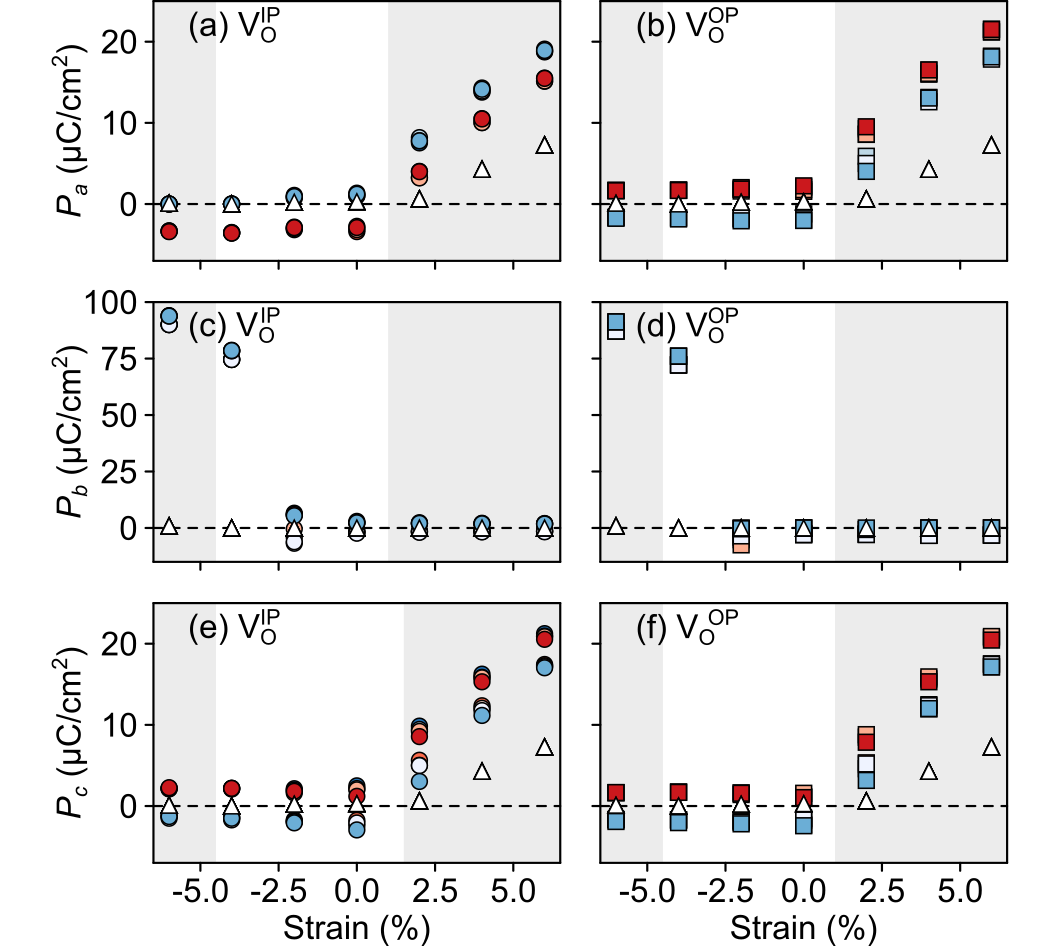}
 \caption{Strain dependence of the polarization component along the (a-b) $a$-, (c-d) $b$-, and (e-f) $c$-axis for the different defect-pair configurations in AFM SMO. (a), (c), and (e) for \ch{V_{Sr}-V_O^{IP}} and (b), (d), and (f) for \ch{V_{Sr}-V_O^{OP}} defects. See Fig.~\ref{fig:confs} for the color code. The shaded grey areas indicate strain ranges with unstable polar modes in stoichiometric SMO and the white triangles correspond to the polarization in stoichiometric SMO.}
\label{fig:pol_AFM_unstrained}
\end{figure}
\begin{figure}
 \centering
 \includegraphics[width=\columnwidth]{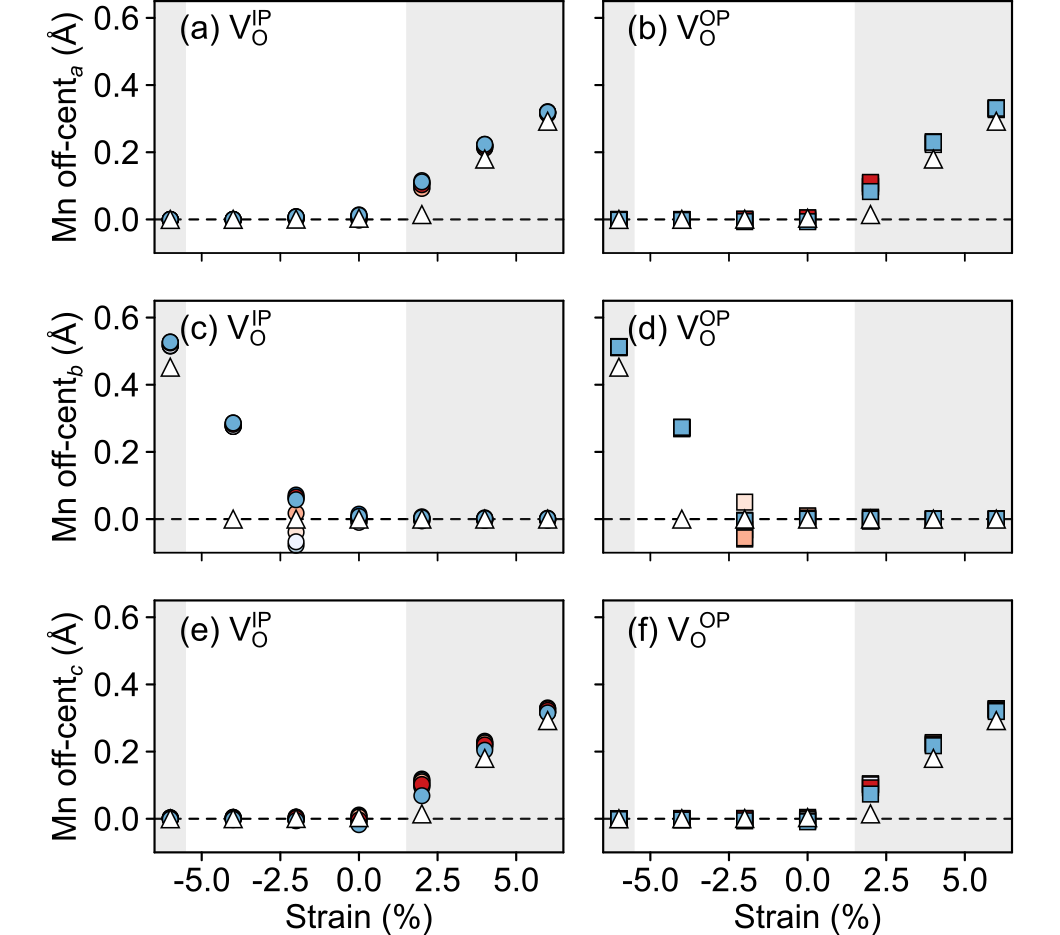}
 \caption{Strain dependence of the average Mn off-centerings along the (a-b) $a$-, (c-d) $b$-, and (e-f) $c$-axis for the different defect-pair configurations in AFM SMO. (a), (c), and (e) for \ch{V_{Sr}-V_O^{IP}} and (b), (d), and (f) for \ch{V_{Sr}-V_O^{OP}} defects. Off-centerings were computed excluding the Mn adjacent to \ch{V_O} to avoid artifacts due to the relaxations of undercoordinated sites. See Fig.~\ref{fig:confs} for the color code. The shaded grey areas indicate strain ranges with unstable polar modes in stoichiometric SMO and the white triangles correspond to the Mn off-centerings in stoichiometric SMO.}
\label{fig:Mnoff_AFM_strain_noquanta}
\end{figure}
In the AFM phase, the components of the polarization in the strained $ac$ plane ($\vec{P}_a$ and $\vec{P}_c$ in Fig.~\ref{fig:pol_AFM_unstrained}) increase steadily with tensile strain, reaching about 20~$\mu$C/cm$^2$ at 6\% strain, in line with the softening of the IP polar modes in the stoichiometric structure (cf. Fig~\ref{fig:Phonons}). We note here that this polarization is of similar magnitude as in conventional ferroelectrics such as \ch{BaTiO3} (22~$\mu$C/cm$^2$)~\cite{Ricca2020LMO, Ederer2005}. This increased polarization is accompanied by an average increase of the Mn off-centering up to about 0.3~\AA\ (see Fig.~\ref{fig:Mnoff_AFM_strain_noquanta}). The larger Mn displacements computed for the defective case with respect to the stoichiometric case (white triangles in Fig.~\ref{fig:Mnoff_AFM_strain_noquanta}), confirm the ability of defect pairs to enhance the polarization. Conversely, compressive strain results in an OP polarization ($\vec{P}_b$) and increased Mn off-centering along the $b$-axis already for about -4\% strain, which is below the critical strain to induce ferroelectricity in stoichiometric SMO (indicated by the gray background shade). The polar defect pairs can hence trigger the ferroelectric phase transition at lower strains and enhance the polarization and Mn off-centering up to 90~$\mu C/cm^2$ and 0.6~\AA\ at -6\%, respectively.

Interestingly, the FM phase shows a different behavior, the computed polarization for all defect configurations being almost constant and close to the polarization in the unstrained structure (see SI\dag\ Figs.~\ref{fig:pol_FM_unstrained}). Only for large compressive strain, when the OP polar modes becomes unstable, an increase of $\vec{P}_b$ is observed. The different behavior of this magnetic order can be explained considering both the strain-dependence of the polar modes and the larger electronic screening in this metallic phase.

%%%%%%%%%%%%%%%%%%%%%%%%%%%%%%%%%%%%%%%%%%%%%%%%%%%%%%%%%%%%%%%%%%%%%%%%%%%%%%%%%%%%%%%
\subsubsection{Ferroelectricity: defect coupling and polarization switching}
%%%%%%%%%%%%%%%%%%%%%%%%%%%%%%%%%%%%%%%%%%%%%%%%%%%%%%%%%%%%%%%%%%%%%%%%%%%%%%%%%%%%%%%

Results discussed so far clearly indicate that \ch{V_{Sr}-V_{O}} defects can be an important source of local polarization in non-polar SMO. Furthermore, different energetically nearly degenerate configurations (see Sec.~\ref{sec:ef_confs}) suggest the possibility of a switchable defect polarization, which could lead to defect-induced ferroelectricity. This effect will however require coupling of the defect dipoles as well as switching of the defect dipoles, which we will investigate in this section.

\begin{figure}
 \centering
 \includegraphics[width=\columnwidth]{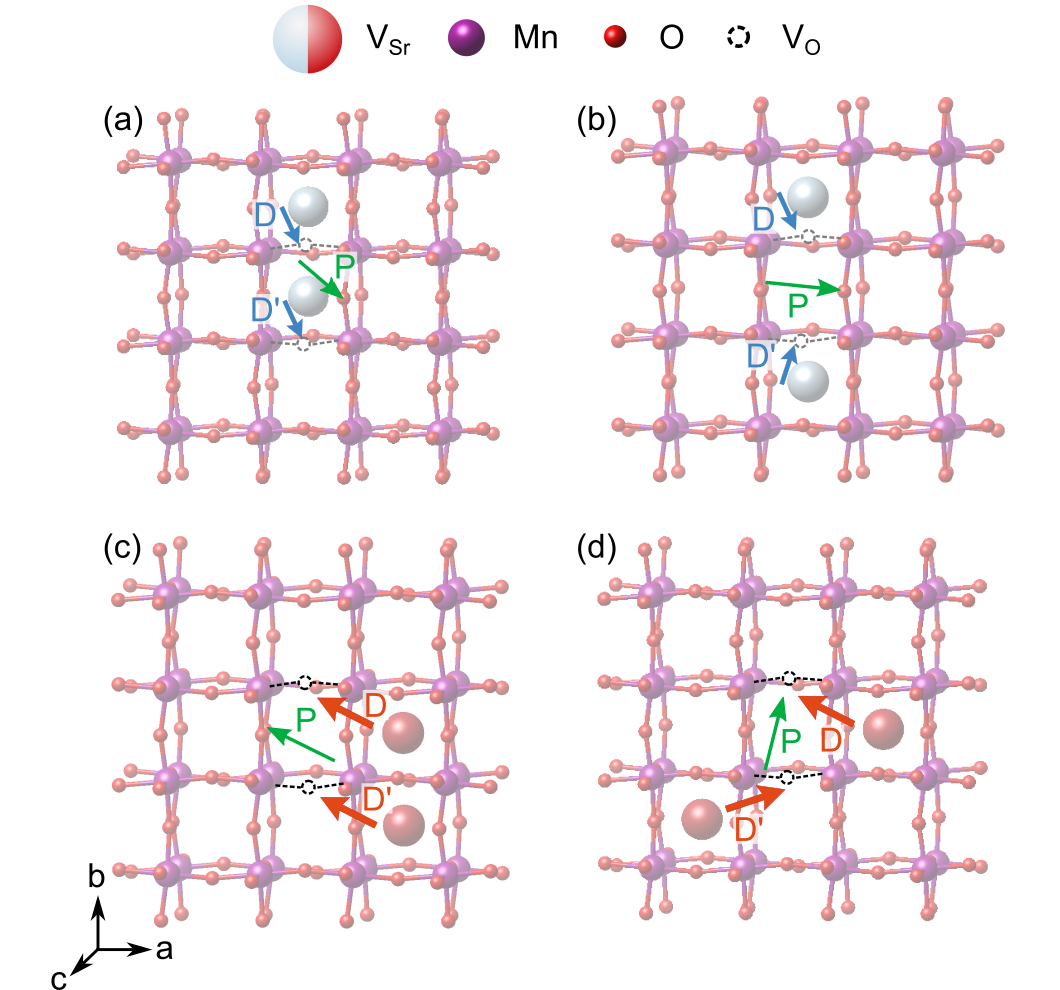}
 \caption{SMO supercell containing two parallel (a and c) or anti-parallel (b and d) defect pairs for (a-b) \ch{V_{Sr}^{NN}-V_O^{IP}} or (c-d) \ch{V_{Sr}^{NNN}-V_O^{IP}} defect pairs. Green, blue and red arrows indicate the direction of polarization ($\vec{P}$) and of the two defect dipoles $\vec{D}$ and $\vec{D'}$ respectively.}
\label{fig:doubledefect}
\end{figure}
When two of the most stable \ch{V_{Sr}^{NN}-V_O^{IP}} vacancy pairs are created in unstrained SMO (Fig. \ref{fig:doubledefect}a and b), the parallel arrangement of their defect dipoles is energetically favored by about 0.13~eV compared to the anti-parallel arrangement. For \ch{V_{Sr}^{NNN}-V_O^{IP}} defect pairs (Fig. \ref{fig:doubledefect}c and d) with larger separation the parallel arrangement is still favored by 0.03~eV. This suggests that coupling of nearby defect-pair dipoles is possible, even at room temperature. The polarization induced by the two defect pairs can be rationalized from the orientation of the defect dipoles: for example, for two parallel \ch{V_{Sr}^{NN}-V_O^{IP}} pairs, the polarization is enhanced mainly along $-b$ and $+c$ compared to a single defect pair, while the anti-parallel arrangement results in an enhancement of $\vec{P}$ along $c$, but in quenching of the polarization along $b$, in line with the opposite orientation of the two dipoles along this axis (see Fig.~\ref{fig:doubledefect}a-b).

\begin{figure}[h]
 \centering
 \includegraphics[width=\columnwidth]{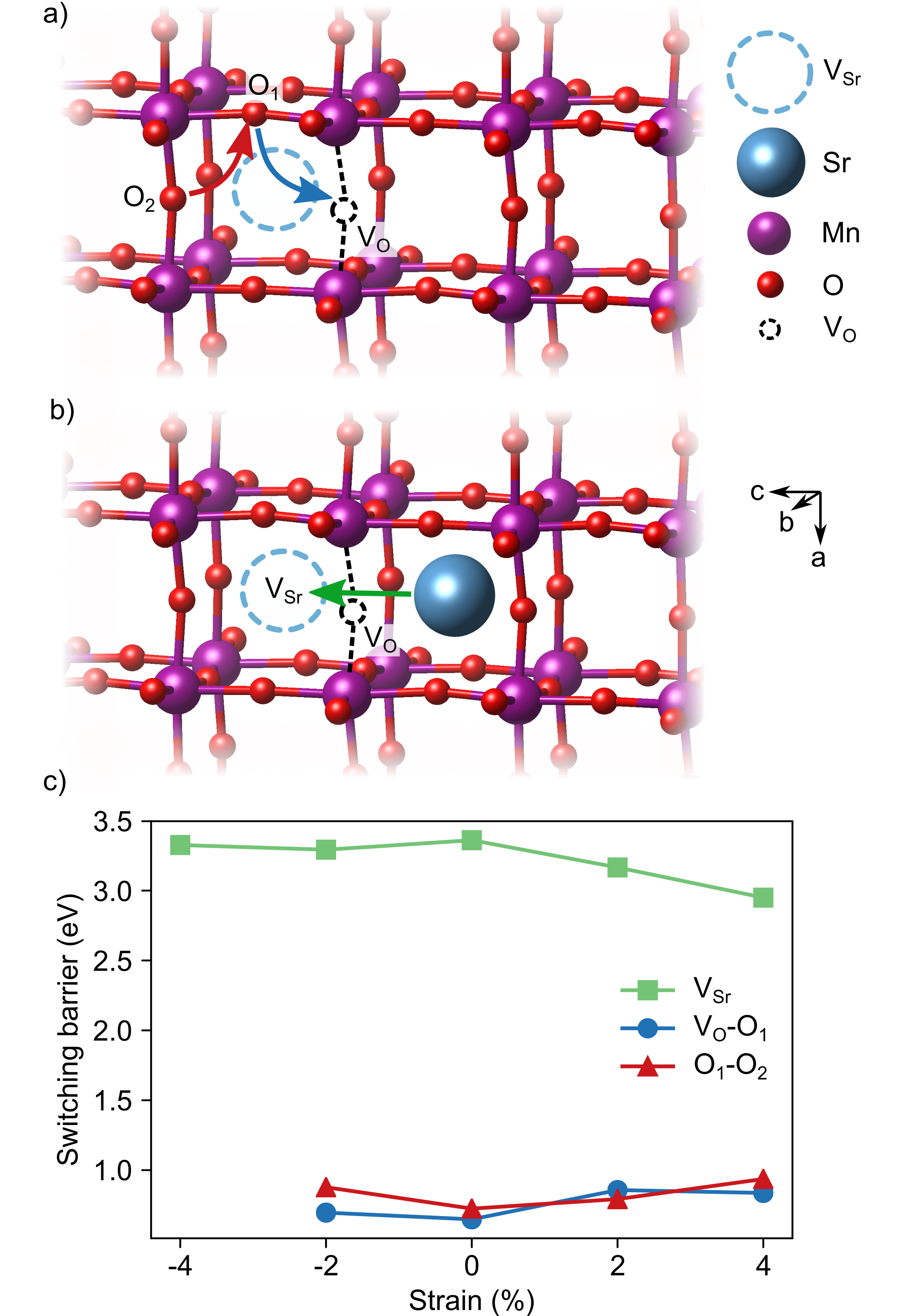}
 \caption{Schematic representation of the a) concerted O migration and b) the Sr migration pathways that invert the defect dipole. c) Evolution of the switching barriers with strain.}
\label{fig:switching}
\end{figure}
Switching of the defect-pair dipoles is the second requirement for defect-induced ferroelectricity in SMO. In presence of \ch{V_{Sr}-V_{O}} defect pairs, switching could take place \textit{via} diffusion of oxygen or strontium vacancies. In the first case, a concerted double jump would move atom \ch{O1} in Fig.~\ref{fig:switching}a into the initial vacancy site, while atom \ch{O2} fills the now vacant \ch{O1} site. The resulting position of the \ch{V_O} at location \ch{O2} inverts the orientation of the defect dipole compared to the initial \ch{V_O} position. For polarization switching \textit{via} Sr diffusion the Sr would follow an approximately linear path between two neighboring Sr sites, as shown in Fig.~\ref{fig:switching}b. In unstrained SMO, the switching barriers are about 0.65-0.72~eV and 3.33~eV for the O and Sr migration, respectively (see Fig.~\ref{fig:switching}c), in good agreement with previous theoretical results in similar perovskite oxides~\cite{Cuong2007, Walsh2011, Klyukin2017}. Even though tensile strain enhances Sr mobility, likely \textit{via} opening the diffusion pathway along the a direction\cite{Kushima:2010dj, Yildiz:2014dy}, O diffusion remains the main pathway for switching the defect dipole. While the barriers for this latter process are larger than the double-well barriers of 0.1~eV in ferroelectric \ch{PbTiO3}~\cite{cohen1992origin}, they are still low enough for polarization reversal \textit{via} electric fields.

%%%%%%%%%%%%%%%%%%%%%%%%%%%%%%%%%%%%%%%%%%%%%%%%%%%%%%%%%%%%%%%%%%%%%%%%%%%%%%%%%%%%%%%
\section{Conclusions}
%%%%%%%%%%%%%%%%%%%%%%%%%%%%%%%%%%%%%%%%%%%%%%%%%%%%%%%%%%%%%%%%%%%%%%%%%%%%%%%%%%%%%%%

In the present work we studied the formation of \ch{V_{Sr}-V_{O}} defect pairs and their impact on the ferroelectricity of \ch{SrMnO3} thin films using DFT+$U$. Our results suggest that polar defect pairs made by Sr cation and O anion divacancies induce defect-pair dipoles from the negatively charged \ch{V_{Sr}} to the positively charged \ch{V_O}, which are an important source of local polarization in non-polar SMO. Electronic screening in the metallic FM phase suppresses significant coupling of these defect-pair dipoles with the lattice polarization. In the semiconducting AFM phase we predict an alignment of the lattice polarization with the defect-pair dipole and, within a sphere of radius 6~\AA\ round the vacancy pairs, an enhanced off-centering of the Mn ions from their high symmetry position in the oxygen octahedra. Divacancies couple with epitaxial strain, which affects their formation energy, allowing for defect ordering, as well as enhancing the polarization in thin films where strain alone could not stabilize a ferroelectric phase. In particular under compressive strain, out-of-plane polarization emerges at significantly lower critical strain in presence of \ch{V_{Sr}-V_{O}} than in the stoichiometric material. Since the direction of these defect-pair dipoles is switchable by an applied electric field and given the tendency of defect-pair dipoles to couple at sufficiently high concentrations, our findings motivate the exploration of intrinsic doping as a parameter to control the ferroelectric transition in complex transition metal oxides.

%%%%%%%%%%%%%%%%%%%%%%%%%%%%%%%%%%%%%%%%%%%%%%%%%%%%%%%%%%%%%%%%%%%%%%%%%%%%%%%%%%%%%%%
\section*{Acknowledgments}
%%%%%%%%%%%%%%%%%%%%%%%%%%%%%%%%%%%%%%%%%%%%%%%%%%%%%%%%%%%%%%%%%%%%%%%%%%%%%%%%%%%%%%%

This research was supported by the NCCR MARVEL, funded by the Swiss National Science Foundation. Computational resources were provided by the University of Bern (on the HPC cluster UBELIX, http://www.id.unibe.ch/hpc) and by the Swiss National Supercomputing Center (CSCS) under project ID mr26.

%%%%%%%%%%%%%%%%%%%%%%%%%%%%%%%%%%%%%%%%%%%%%%%%%%%%%%%%%%%%%%%%%%%%%%%%%%%%%%%%%%%%%%%
\bibliography{references}
%%%%%%%%%%%%%%%%%%%%%%%%%%%%%%%%%%%%%%%%%%%%%%%%%%%%%%%%%%%%%%%%%%%%%%%%%%%%%%%%%%%%%%%

%%%%%%%%%%%%%%%%%%%%%%%%%%%%%%%%%%%%%%%%%%%%%%%%%%%%%%%%%%%%%%%%%%%
%%%   SUPPLEMENTARY   %%%
%%%%%%%%%%%%%%%%%%%%%%%%%%%%%%%%%%%%%%%%%%%%%%%%%%%%%%%%%%%%%%%%%%%

%reset all style and numbering
\clearpage
\clearpage %needed for two-page reference section
\setcounter{page}{1}
\renewcommand{\thetable}{S\arabic{table}} 
\setcounter{table}{0}
\renewcommand{\thefigure}{S\arabic{figure}}
\setcounter{figure}{0}
\renewcommand{\thesection}{S\arabic{section}}
\setcounter{section}{0}
\renewcommand{\theequation}{S\arabic{equation}}
\setcounter{equation}{0}
\onecolumngrid

%create title
\begin{center}
\textbf{Supplementary information for\\\vspace{0.5 cm}
\large Ferroelectricity promoted by cation/anion divacancies in \ch{SrMnO3}\\\vspace{0.3 cm}}
Chiara Ricca,$^{1, 2}$, Danielle Berkowitz$^{1}$ and Ulrich Aschauer$^{1, 2}$

\small
$^1$\textit{Department of Chemistry and Biochemistry, University of Bern, Freiestrasse 3, CH-3012 Bern, Switzerland}

$^2$\textit{National Centre for Computational Design and Discovery of Novel Materials (MARVEL), Switzerland}

(Dated: \today)
\end{center}

%%%%%%%%%%%%%%%%%%%%%%%%%%%%%%%%%%%%%%%%%%%%%%%%%%%%%%%%%%%%%%%%%%%%%%%%%%%%%%%%%%%%%%%
\section{\label{sec:si_eg} Strain-dependence of SMO band-gap}
%%%%%%%%%%%%%%%%%%%%%%%%%%%%%%%%%%%%%%%%%%%%%%%%%%%%%%%%%%%%%%%%%%%%%%%%%%%%%%%%%%%%%%%

\begin{figure}[h]
 \centering
 \includegraphics[width=0.4\textwidth]{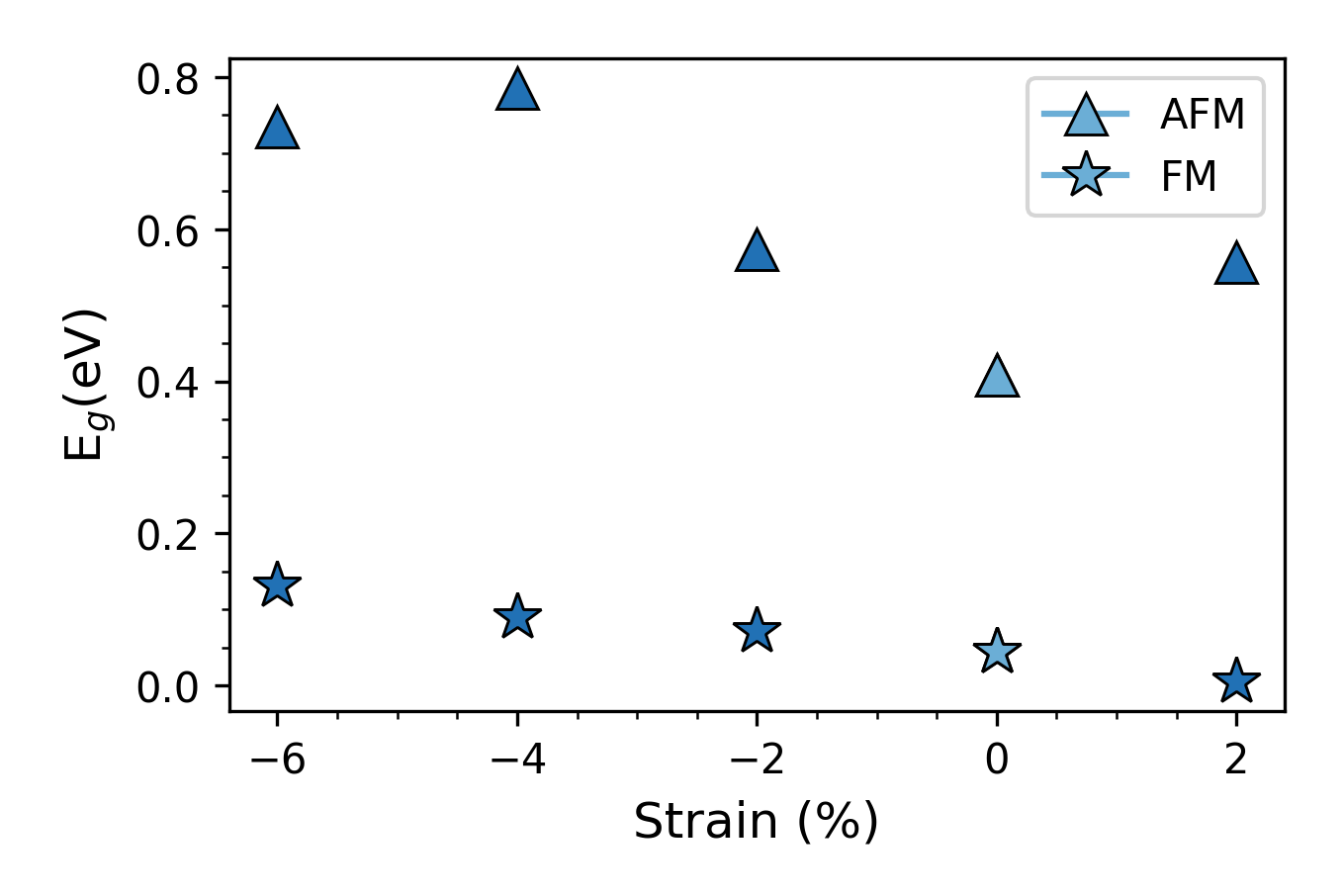}
 \caption{Evolution of the SMO bandgap ($E_\mathrm{g}$) in the AFM and FM phase as a function of biaxial strain.}
 \label{fig:eg_vs_strain}
\end{figure}

For the metallic FM phase, compressive strain results in a small band gap ($E_\textrm{g}$) almost linearly increasing with increasing strain (see Fig.~\ref{fig:eg_vs_strain}). Instead, for the AFM phase, $E_\textrm{g}$ increases up to -4\% and decreases for larger strains, which strongly destabilizes this magnetic order.
%

%%%%%%%%%%%%%%%%%%%%%%%%%%%%%%%%%%%%%%%%%%%%%%%%%%%%%%%%%%%%%%%%%%%%%%%%%%%%%%%%%%%%%%%
\section{\label{sec:si_pdos} Electronic Properties for \ch{V_{Sr}-V_{O}} defect pairs in SMO}
%%%%%%%%%%%%%%%%%%%%%%%%%%%%%%%%%%%%%%%%%%%%%%%%%%%%%%%%%%%%%%%%%%%%%%%%%%%%%%%%%%%%%%%
%
\begin{figure}[h]
 \centering
 \includegraphics{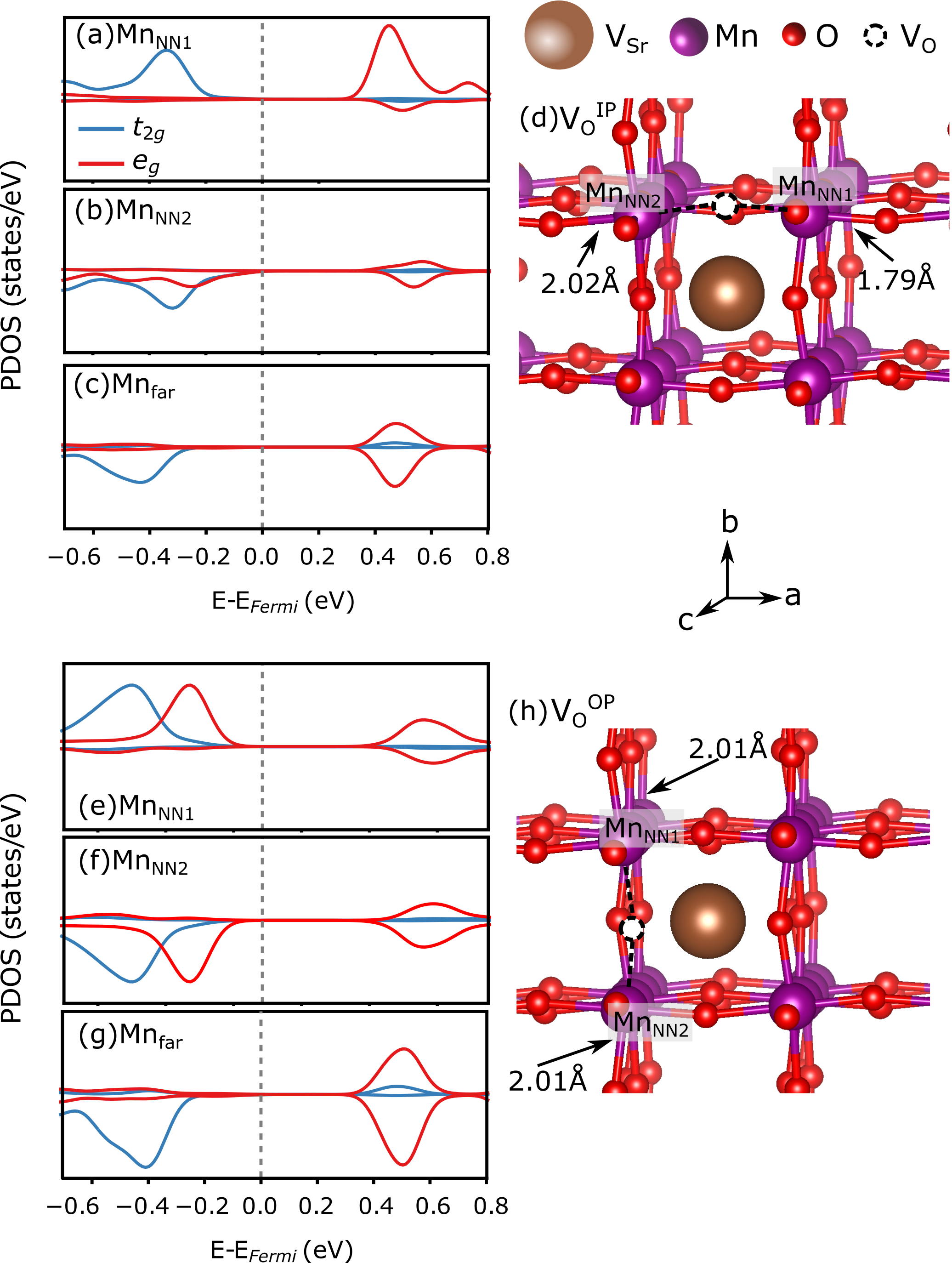}
 \caption{Atom- and orbital-projected density of states (PDOS) for the Mn-$3d$ states of the Mn sites in nearest-neighbor (NN) positions to \ch{V_O} (a,b,e,f) and for a Mn atom far from the defect (c and g) for a \ch{V_{Sr}-V_{O}^{IP}} (a-c) or a \ch{V_{Sr}-V_{O}^{OP}} (e-g) defect pair. The vertical dashed line indicates the Fermi level. Panels d) and h) shows the most relevant \ch{Mn-O} bond changes taking place on the NN Mn atoms after \ch{V_{Sr}-V_{O}^{IP}} and \ch{V_{Sr}-V_{O}^{OP}} formation, respectively.}
 \label{fig:PDOS}
\end{figure}
Due to the formal charge compensation of the \ch{V_{Sr}^{''}} and \ch{V_{O}^{..}} defects, the formation of a neutral \ch{V_{Sr}-V_{O}} defect pair should not be associated with any excess charge (resulting, for example, in the reduction of Mn sites). However, the density of states reported in Fig.~\ref{fig:PDOS} for a \ch{V_{Sr}-V_{O}^{IP}} and a \ch{V_{Sr}-V_{O}^{OP}} suggests a partial reduction of one and two, respectively, Mn atoms adjacent to \ch{V_O} forming \ch{Mn^{($3+\delta$)+}} ions as a consequence of the structural relaxations taking place upon \ch{V_O} formation. In particular, in the case of \ch{V_{O}^{OP}}, after breaking the two equivalent \ch{Mn-O-Mn} bonds along the $b$-axis, the remaining \ch{Mn-O} bonds along this direction for the Mn in nearest neighbor position to the O vacancy are elongated by about 0.10~\AA\ and reach 2.01~\AA. This lowers the energy of the $e_g$ states of these sites, which lie at the bottom of SMO conduction band. The lowering of the energy of these $e_g$ orbitals to just below the Fermi energy results in partial filling of these states and consequently in a partial Mn reduction. This, however, is a consequence of the underestimation of the SMO band gap in DFT+$U$ calculations~\citeSI{Ricca2019SI}. Similar arguments apply to \ch{V_{Sr}-V_{O}^{IP}}, but in this case only one of the two Mn ions adjacent to the removed O atom has the remaining \ch{Mn-O} bond elongated by about 0.05~\AA\ to reach 2.01~\AA\ along the \ch{Mn-V_O-Mn} axis. For the other site adjacent to the oxygen vacancy the remaining \ch{Mn-O} bond is shortened by 0.13~\AA\ to a relaxed length of 1.79~\AA. As a result, only one Mn is artificially reduced.

\clearpage
\newpage
%%%%%%%%%%%%%%%%%%%%%%%%%%%%%%%%%%%%%%%%%%%%%%%%%%%%%%%%%%%%%%%%%%%%%%%%%%%%%%%%%%%%%%%
\section{\label{sec:si_Mnrelaxations} Structural relaxations and {Mn} off-centerings}
\begin{figure}[h]
 \centering
 \includegraphics[width=0.9\textwidth]{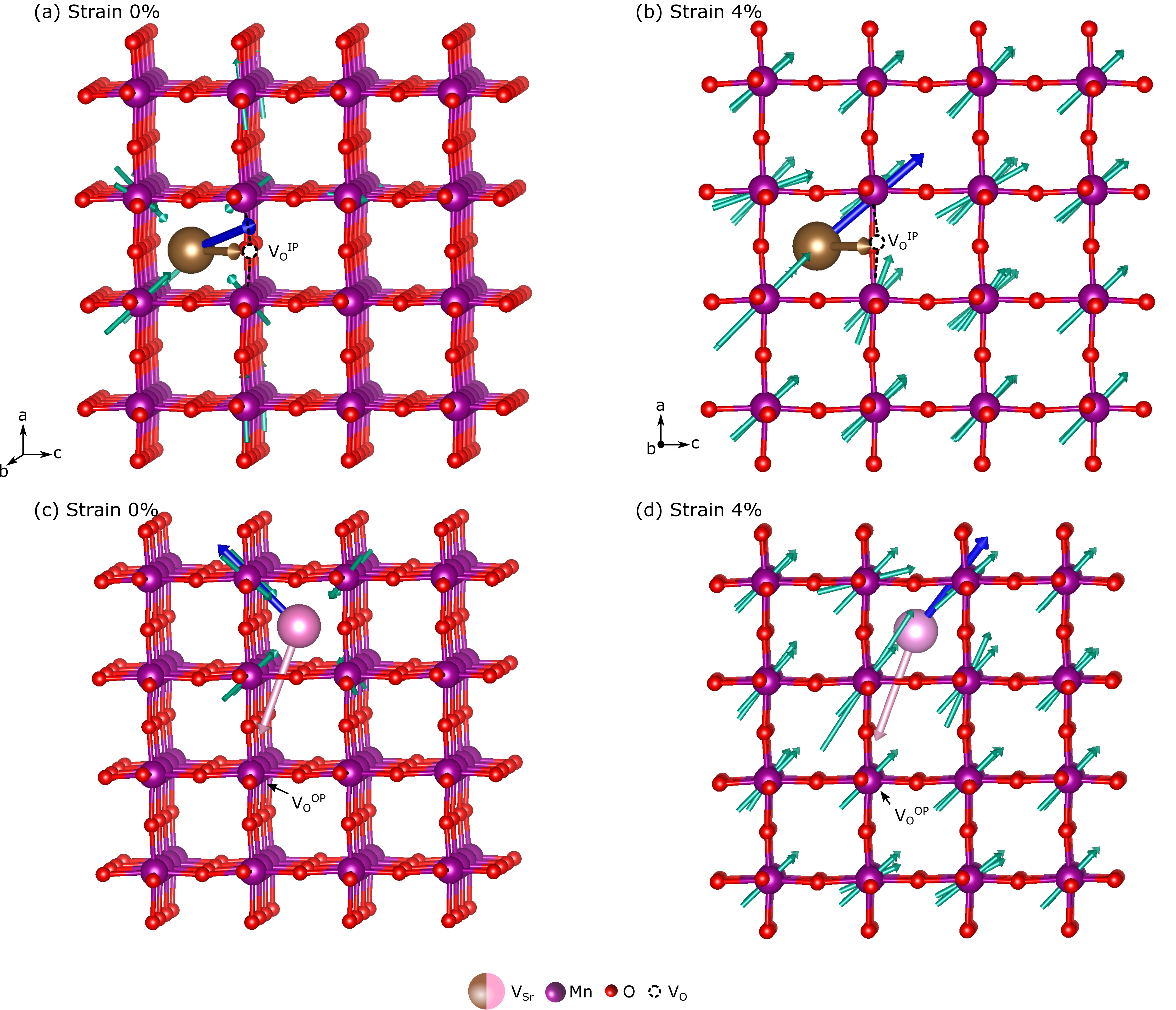}
 \caption{Off-centerings of the Mn atoms upon formation of a-b) one \ch{V_{Sr}-V_{O}^{IP}} and c-d) one \ch{V_{Sr}-V_{O}^{OP}} in unstrained SMO (plots a) and c) and for 4\% tensile strain (plots b) and d). The green arrows indicate the vector corresponding to the off-centering of each Mn atom, while the brown/pink and blue arrows correspond to the defect dipole and polarization vectors, respectively.}
 \label{fig:Mnrelaxations}
\end{figure}

%%%%%%%%%%%%%%%%%%%%%%%%%%%%%%%%%%%%%%%%%%%%%%%%%%%%%%%%%%%%%%%%%%%%%%%%%%%%%%%%%%%%%%%
Structural relaxations upon formation of a \ch{V_{Sr}-V_{O}} defect pair in the unstrained SMO structure mainly result in off-centering of Mn atoms in nearest neighbor (NN) positions to \ch{V_{Sr}} from their high symmetry positions and towards the cation vacancy (see Figs.~ \ref{fig:Mnrelaxations}a and c). When \ch{V_{Sr}} is NN to \ch{V_{O}}, the Mn adjacent to both vacancies cannot move towards \ch{V_{Sr}} because of the broken \ch{Mn-O-Mn} bond. Overall the defect polarizes the surrounding atoms resulting in a polarization vector mainly aligned with the defect dipole, but forming a small angle with it as a consequence of the displacements of the Mn atoms in next nearest neighbor (NNN) positions to \ch{V_O} along the axis where the \ch{Mn-O-Mn} bonds is broken. Instead, for configurations where \ch{V_{Sr}} lies at distances larger than 6~\AA\ from \ch{V_O}, a full contraction of the lattice, due to the displacements of all the Mn sites adjacent to \ch{V_{Sr}} towards the vacancy, results in the polarization not being aligned with the defect dipole. For a larger tensile strain of 4\%, when the in-plane polar modes are unstable, larger off-centering for all the Mn atoms are observed along the [101] direction: this effect partially enhances or counteracts the relaxations of Mn adjacent to \ch{V_{Sr}} when the displacement due to the cation vacancy formation is aligned or opposite to the in-plane polarization direction (see Figs.~ \ref{fig:Mnrelaxations}b and d).

\newpage
%%%%%%%%%%%%%%%%%%%%%%%%%%%%%%%%%%%%%%%%%%%%%%%%%%%%%%%%%%%%%%%%%%%%%%%%%%%%%%%%%%%%%%%
\section{\label{sec:si_polFM} Polarization in the FM order}
%%%%%%%%%%%%%%%%%%%%%%%%%%%%%%%%%%%%%%%%%%%%%%%%%%%%%%%%%%%%%%%%%%%%%%%%%%%%%%%%%%%%%%%

\begin{figure}[h]
 \centering
 \includegraphics[width=0.5\textwidth]{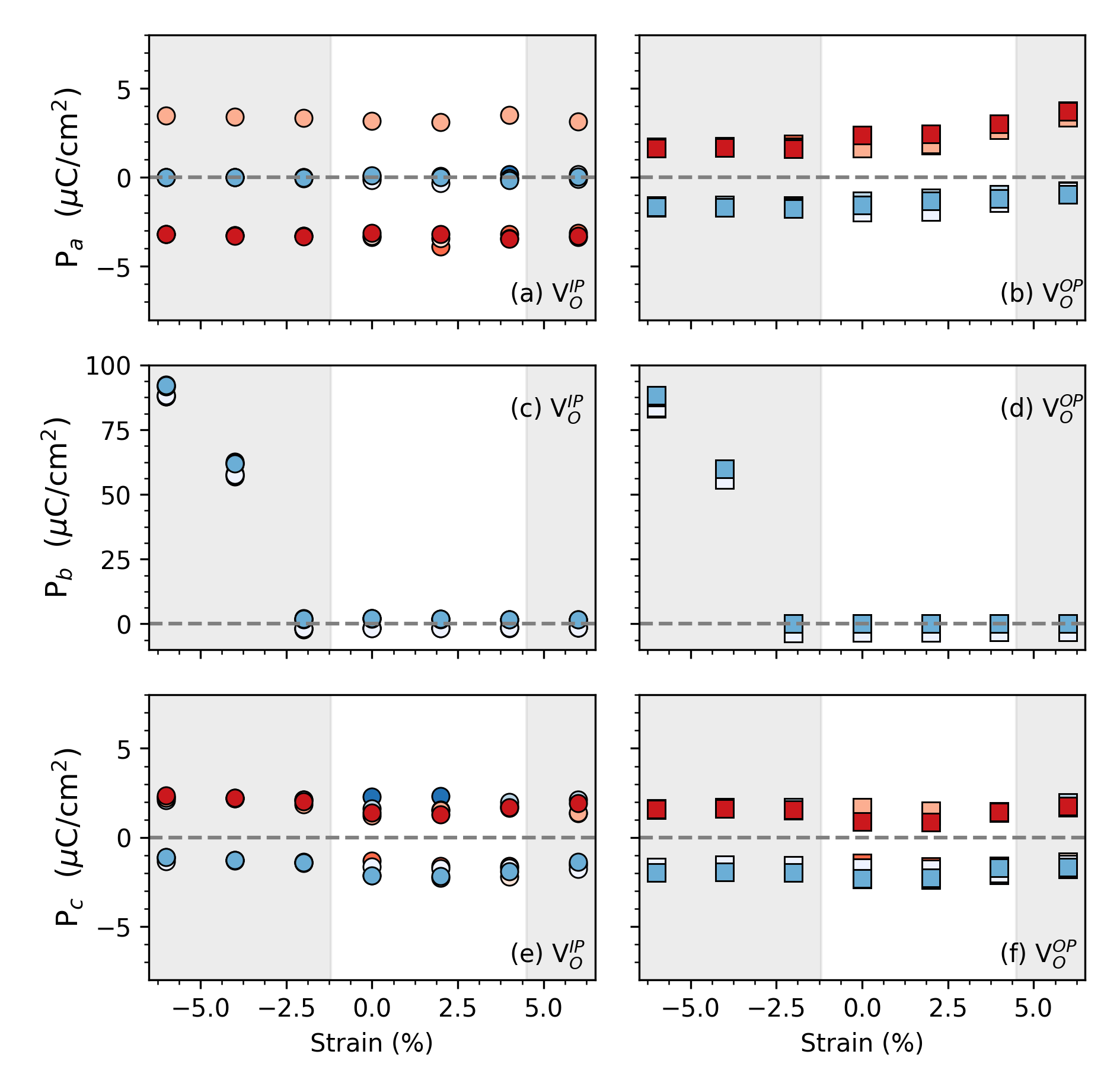}
 \caption{Strain dependence of the component of the polarization along the (a-b) $a$-, (c-d) $b$-, and (e-f) $c$-axis for the different defect-pair configurations in the FM phase of SMO. (a), (c), and (e) plots for \ch{V_{Sr}-V_O^{IP}} and (b), (d), and (f) plots for \ch{V_{Sr}-V_O^{OP}} defects. See Fig.~\ref{fig:confs} in the main text for the color code. The shaded grey areas indicate strain ranges with unstable polar modes in stoichiometric SMO.}
\label{fig:pol_FM_unstrained}
\end{figure}

The strain dependent polarization of SMO with \ch{V_{Sr}-V_{O}} defect pairs is the result of a complex interplay between defect chemistry and electronic and magnetic degrees of freedom. While for the insulating AFM phase, polarization can be enhanced by tensile strain, in the metallic FM phase the polarization is fairly constant with respect to the applied epitaxial strain in the range between 0 and 6\% (see Fig.~\ref{fig:pol_FM_unstrained}). Interestingly, compressive strain beyond -2\% can result in an enhancement of the out-of-plane component of the polarization, in line with the larger instability of the OP polar modes observed already in the stoichiometric FM phase (cf. Fig.~\ref{fig:Phonons} in the main text). In summary, the behavior of the FM phase can be rationalized considering both the larger electronic screening of the defect dipole in the metallic FM order, which results in a lower sensitivity of the polarization of the defective cells to the applied strain, and the strain-dependence of the polar modes frequencies in the stoichiometric FM SMO.
%

%%%REFERENCES%%%
%\clearpage
\bibliographystyleSI{apsrev4-1}
\bibliographySI{references}

\end{document}